\documentclass[american,english]{IEEEtran}
\usepackage[T1]{fontenc}
\usepackage[latin9]{inputenc}
\usepackage{color}
\usepackage{float}
\usepackage{textcomp}
\usepackage{amsthm}
\usepackage{amsmath}
\usepackage{amssymb}
\usepackage{stackrel}
\usepackage{graphicx}
\usepackage{esint}

\makeatletter

\floatstyle{ruled}
\newfloat{algorithm}{tbp}{loa}
\providecommand{\algorithmname}{Algorithm}
\floatname{algorithm}{\protect\algorithmname}

\theoremstyle{plain}
\newtheorem{thm}{\protect\theoremname}

\usepackage{colortbl}
\usepackage{cite}
\usepackage{nomencl}
\usepackage{flushend}
\usepackage{stfloats}
\usepackage[margin=8pt,font=footnotesize,labelfont=md,labelsep=colon]{caption}

\@ifundefined{showcaptionsetup}{}{%
 \PassOptionsToPackage{caption=false}{subfig}}
\usepackage{subfig}
\makeatother

\usepackage{babel}
\addto\captionsamerican{\renewcommand{\algorithmname}{Algorithm}}
\addto\captionsamerican{\renewcommand{\theoremname}{Theorem}}
\addto\captionsenglish{\renewcommand{\theoremname}{Theorem}}
\providecommand{\theoremname}{Theorem}

\begin{document}

\title{Sum Rate and Fairness Analysis for the MU-MIMO Downlink under PSK
Signalling: Interference Suppression vs Exploitation}

\author{Abdelhamid Salem, \textit{\normalsize{}Student Member, IEEE}{\normalsize{},}
Christos Masouros, \textit{\normalsize{}Senior Member, IEEE}, and
Kai-Kit Wong, \textit{\normalsize{}Fellow, IEEE}\\
\thanks{The authors are with the department of Electronic and Electrical Engineering,
University College London, London, UK, (emails: \{a.salem, c.masouros,
kai-kit.wong\}@ucl.ac.uk). Part of this paper has been submitted to
IEEE WCNC\textquoteright 19.%
} }
\maketitle
\begin{abstract}
In this paper, we analyze the sum rate performance of multi-user multiple-input
multiple-output (MU-MIMO) systems, with a finite constellation phase-shift
keying (PSK) input alphabet. We analytically calculate and compare
the achievable sum rate in three downlink transmission scenarios:
1) without precoding, 2) with zero forcing (ZF) precoding 3) with
closed form constructive interference (CI) precoding technique. In
light of this, new analytical expressions for the average sum rate
are derived in the three cases, and Monte Carlo simulations are provided
throughout to validate the analysis. Furthermore, based on the derived
expressions, a power allocation scheme that can ensure fairness among
the users is also proposed. \textcolor{blue}{ }The results in this
work demonstrate that, the CI strictly outperforms the other two schemes,
and the performance gap between the considered schemes increases with
increase in the MIMO size. In addition, the CI provides higher fairness
and the power allocation algorithm proposed in this paper can achieve
maximum fairness index. \end{abstract}


\section{Introduction}

The recent decades have witnessed the widespread application of multi-user
multiple-input multiple-output (MU-MIMO) communication systems, due
to their high spectral efficiency and reliability \cite{MIMO1,MIMO2,MIMO3}.
However, these potential advantages of MU-MIMO systems are often undermined
by strong interference in practical wireless systems \cite{MIMO1,MIMO2,MIMO3}.
Consequently, considerable amount of researches have focused on reducing/
mitigating the interference in MU-MIMO channels \cite{MIMO3,Twway,interference1}. 

A number of information theoretic works have studied the sum rate
performance of MU-MIMO systems by assuming Gaussian input signals.
However, in practical communication systems, signals are generated
from finite discrete constellation sets. In light of this, several
works have considered MU-MIMO systems for finite alphabet input signals.
In \cite{MIMO3} optimal linear precoder for MU-MIMO interference
channels with finite alphabet inputs was designed. In \cite{interference1}
the design of linear precoders in multi-cell MIMO systems for finite
alphabet signals was studied. The authors in \cite{MIMOPSK1} considered
the capacity of a MIMO fading channel with PSK input alphabet; in
this work a downlink transmission without precoding was studied. In
\cite{precoding1} the design of optimal precoders which maximize
the mutual information of MIMO channels were investigated, by assuming
non-Gaussian inputs of finite alphabet. The design of linear transmit
precoder for MIMO broadcast channels with finite alphabet input signals
was investigated in \cite{precoding2}, where an explicit expression
for the achievable rate region was derived. The work in \cite{precoding3}
proposed low complexity precoding scheme that aimed to maximize the
mutual information for MIMO systems with finite alphabet inputs. Linear
precoder design that maximizes the average mutual information of MIMO
fading channels with finite-alphabet inputs was proposed in \cite{precoding4},
in which the statistical channel state information (CSI) was assumed
to be known at the transmitter side. In \cite{precoding5}, the authors
studied a linear precoding for MIMO channels with finite discrete
inputs, in which the capacity region for the MU-MIMO has been derived.
In \cite{precoding6} a linear precoder design for MU-MIMO transceivers
under finite alphabet inputs was proposed, where the optimal transmission
strategies in both low and high signal-to-noise ratio (SNR) regions
were studied. Although the aforementioned algorithms produced optimal
performances, the fact that they have no closed form solutions and
their resulting high computational complexity make them inapplicable
in practical scenarios.

Recently, constructive interference (CI) precoding technique has been
proposed to enhance the performance of downlink MU-MIMO systems \cite{CI1,A,CI2,CI3}.
In contrast to the conventional techniques where the knowledge of
the interference is used to cancel it, the main idea of the CI is
to use the interference to improve the system performance. Specifically,
the CI precoding technique exploits interference that can be known
to the transmitter to increase the useful signal received power \cite{CI1,A,CI2,CI3}.
That is, with the knowledge of the users\textquoteright{} data symbols
and CSI, the interference can be classified as constructive and destructive.
The interference signal is considered to be constructive to the transmitted
signal if it moves the received symbols away from the decision thresholds
of the constellation towards the direction of the desired symbol.
Accordingly, the transmit precoding can be designed such that the
resulting interference is constructive to the desired symbol.

The concept of the CI has been extensively studied in literature.
This line of work has been introduced in \cite{CI1}, where the CI
precoding scheme for the downlink of PSK-based MIMO systems has been
proposed. In this work it was shown that the system performance can
be enhanced by exploiting the interference signals. As a result, the
effective signal to interference-plus-noise ratio (SINR) can be enhanced
without the need to increase the transmitted signal power at the base
station (BS). In \cite{A} the concept of CI was used to design an
optimization based precoder in the form of pre-scaling for the first
time. Thereoff, \cite{CI2} proposed transmit beamforming schemes
for the MU-MIMO down-link that minimize the transmit power for generic
PSK signals. In \cite{CI3,CI5} CI precoding scheme was applied in
wireless power transfer scenario in order to minimize the transmit
power while guaranteeing the energy harvesting and the quality of
service (QoS) constraints for PSK messages. Further work in \cite{CI4}
applied the CI concept to massive multi-input multi-output (M-MIMO)
systems. Very recently, the authors in \cite{angLi} derived closed-form
precoding expression for CI exploitation in the MU-MIMO down-link.
The closed-form precoder in this work has for the first time made
the application of CI exploitation practical, and has further paved
the way for the development of communication theoretic analyses of
the benefits of CI, which is the focus of this work.

Accordingly this paper investigates the sum rate of downlink transmission
with finite constellation PSK signalling for interference suppression
and interference exploitation techniques. Within this context, three
transmission techniques are considered. The first is based on scenarios
where the CSI is unknown at the BS, in which we study downlink transmission
without precoding. The other two transmission schemes are based on
full knowledge of the users' CSI at the BS. In the second transmission
scheme zero forcing (ZF) precoding technique is considered, and in
the third scheme closed form CI precoding technique is analyzed for
the first time. In this regard, explicit expressions for the average
sum rate are derived in each transmission scheme. Then, based on the
derived expressions the fairness among the users is also investigated. 

For clarity we list the major contributions of this work as:
\begin{enumerate}
\item Firstly, new explicit analytical expressions for the average achievable
rate upper bound of MU-MIMO with PSK inputs are derived for a) un-precoded
transmission, b) ZF precoded transmission and c) CI precoded transmission,
considering the channels to be of Rayleigh fading. 
\item Based on the above analysis, a power allocation scheme that can provide
fairness among the users is proposed.
\item The impact of the main system parameters on the system performance
of the considered schemes are examined and investigated. 
\end{enumerate}
Results provided in this paper show that CI scheme outperforms the
other two schemes, for the same system parameters. Also, it is shown
that increasing the SNR and the number of the BS antennas enhances
the system performance, whereas the gap between the minimum transmission
power required for ZF and CI to achieve same target rate is almost
fixed with increasing the distance between the BS and the users. Furthermore,
the CI provides higher fairness than ZF and the power allocation algorithm
proposed in this paper achieves high fairness index. 

Next, Section \ref{sec:System-Model} describes the system model under
consideration. Sections \ref{sec:Conventional-Transmission}, \ref{sec:Zero-Forcing-Precoding},
and \ref{sec:Constructive-Interference-Precod} derive the analytical
expressions for the average sum rate in conventional transmission,
ZF and CI precoding techniques, respectively. Section \ref{sec:Transmission-Power-Minimization},
considers the users fairness in the three transmission schemes. Numerical
and simulation results are presented and discussed in Section \ref{sec:Numerical-Results}.
Finally, the main conclusions of this work are stated in Section \ref{sec:Conclusions}.

Notations: $h$, $\mathbf{h}$, and \textbf{$\mathbf{H}$} denote
a scalar, a vector and a matrix, respectively. $(\cdot)^{H}$, $(\cdot)^{T}$
and $\textrm{diag}\left(.\right)$ denote conjugate transposition,
transposition and diagonal of a matrix, respectively. $\mathcal{E}\left[.\right]$
denotes average operation. $\left[\mathbf{h}\right]_{k}$ denotes
the $k^{th}$ element in $\mathbf{h}$ , $\left|.\right|$ denotes
the absolute value, , and $\left\Vert .\right\Vert ^{2}$ denotes
the second norm. $\mathbb{C}^{K\times N}$ represents an K \texttimes N
matrix, and $\mathbf{I}$ denotes the identity matrix.

\section{System Model\label{sec:System-Model}}

Consider a downlink MU-MIMO system, consisting of a BS equipped with
$N$ antennas communicating with $K$ single antenna users, where
$N\geq K$. All the channels are modeled as independent identically
distributed (i.i.d) Rayleigh fading channels.\textcolor{black}{{} }The
channel matrix between the BS and the $K$ users is denoted by $\mathbf{H}\in\mathbb{C}^{K\times N}$,
which can be represented as $\mathbf{H}=\mathbf{D}^{1/2}\mathbf{H}_{1}$where
$\mathbf{H}_{1}\in\mathbb{C}^{K\times N}$ contains i.i.d $\mathcal{CN}\left(\text{0,}1\right)$
entries which represent small scale fading coefficients and $\mathbf{D}\in\mathbb{C}^{K\times K}$
is a diagonal matrix with $\left[\mathbf{D}\right]_{kk}=\varpi_{k}$
which represent the path-loss attenuation $\varpi_{k}=d_{k}^{-m}$,
$d_{k}$ is the distance between the BS and the $k^{th}$ user and
$m$ is the path loss exponent. It is also assumed that the signal
is equiprobably drawn from an $M$-PSK constellation and denoted as
$\mathbf{s}\in\mathbb{C}^{K\times1}$ \cite{angLi}. The received
signal at the $k^{th}$ user can be expressed as,

\begin{equation}
y_{k}=\mathbf{h}_{k}\mathbf{W}\mathbf{s}+n_{k}\label{eq:1}
\end{equation}

\noindent where $\mathbf{h}_{k}$ is the channel vector from the BS
to user $k$, $\mathbf{W}$ is the precoding matrix, $n_{k}$ is the
additive wight Gaussian noise (AWGN) at the $k^{th}$ user, $n_{k}\sim\mathcal{CN}\left(\text{0, }\sigma_{k}^{2}\right)$.

It was shown in \cite{MIMOPSK1,precoding6,precoding2} that, the achievable
rate for the $k$-th user in general MU-MIMO system with finite constellation
signaling, is given by,

\[
R_{k}=N\log_{2}M-\frac{1}{M^{N}}\stackrel[m=1]{M^{N}}{\sum}\mathcal{E}_{\mathbf{h,}\mathbf{n}}\left\{ \log_{2}\stackrel[i=1]{M^{N}}{\sum}e^{\frac{-\left\Vert \mathbf{h}_{k}\mathbf{W}\mathbf{\boldsymbol{s}}_{m,i}+n_{k}\right\Vert ^{2}}{\sigma_{k}^{2}}}\right\} 
\]
\begin{equation}
+\frac{1}{M^{N-1}}\stackrel[c=1]{M^{N-1}}{\sum}\mathcal{E}_{\mathbf{h,}\mathbf{n}}\left\{ \log_{2}\stackrel[t=1]{M^{N}}{\sum}e^{\frac{-\left\Vert \mathbf{h}_{k}\mathbf{W}\mathbf{\boldsymbol{s}}_{c,t}+n_{k}\right\Vert ^{2}}{\sigma_{k}^{2}}}\right\} ,
\end{equation}

\noindent where $\mathbf{s}_{m,i}=\mathbf{s}_{m}-\mathbf{s}_{i}$,
$\mathbf{s}_{m}$ and $\mathbf{s}_{i}$ contain symbols taken from
the $M$ signal constellation.

In the following, the average sum rate is derived for three cases,
without precoding, with ZF precoding and with CI precoding technique.

\section{Downlink Transmission Without Precoding\label{sec:Conventional-Transmission}}

In this case the BS transmits the users' signals without any precoding
technique, such scenario occurs when the CSI of the users is unknown
at the BS. Let the $k^{th}$signal be equiprobably drawn from an $M$-PSK
constellation, the average rate at a user $k$ can be written as \cite{interference1,precoding6,precoding2},

\[
\bar{R_{k}}=\log_{2}M^{N}-\underset{T_{1}}{\underbrace{\frac{1}{M^{N}}\stackrel[m=1]{M^{N}}{\sum}\mathcal{E}_{\mathbf{h,}n_{k}}\log_{2}\stackrel[i=1]{M^{N}}{\sum}e^{\frac{-\left|\sqrt{p_{N}}\mathbf{h}_{k}\mathbf{s}_{m,i}+n_{k}\right|^{2}}{\sigma_{k}^{2}}}}}
\]

\begin{equation}
+\underset{T_{2}}{\underbrace{\frac{1}{M^{N-1}}\stackrel[\underset{c\neq k}{c=1}]{M^{N-1}}{\sum}\mathcal{E}_{\mathbf{h,}n_{k}}\log_{2}\stackrel[\underset{t\neq k}{t=1}]{M^{N-1}}{\sum}e^{\frac{-\left|\sqrt{p_{N}}\mathbf{h}_{k}\mathbf{s}_{c,t}+n_{k}\right|^{2}}{\sigma_{k}^{2}}}}},\label{eq:7}
\end{equation}

\noindent where $p_{N}=\frac{p}{N}$, is the power transmitted by
each antenna and $p$ is the total power transmission. The second
and third terms in (\ref{eq:7}), $\left\{ T_{1},T_{2}\right\} $,
can be simplified to more familiar formulas. The second term, $T_{1}$,
can be simplified by taking the $j^{th}$ term, $\left(e^{\frac{-\left|\sqrt{p_{N}}\mathbf{h}_{k}\mathbf{s}_{m,j}+n_{k}\right|^{2}}{\sigma_{k}^{2}}}\right)$,
out as follows

\[
T_{1}=\frac{1}{M^{N}}\stackrel[m=1]{M^{N}}{\sum}\mathcal{E}_{\mathbf{h,}n_{k}}\left\{ \log_{2}e^{\frac{-\left|\sqrt{p_{N}}\mathbf{h}_{k}\mathbf{s}_{m,j}+n_{k}\right|^{2}}{\sigma_{k}^{2}}}\right.
\]

\begin{equation}
\times\left.\left(1+\stackrel[i=1,i\neq j]{M^{N}}{\sum}e^{\frac{-\left|\sqrt{p_{N}}\mathbf{h}_{k}\mathbf{s}_{m,i}+n_{k}\right|^{2}+\left|\sqrt{p_{N}}\mathbf{h}_{k}\mathbf{s}_{m,j}+n_{k}\right|^{2}}{\sigma_{k}^{2}}}\right)\right\} 
\end{equation}

\[
=\frac{1}{M^{N}}\stackrel[m=1]{M^{N}}{\sum}\mathcal{E}_{\mathbf{h,}n_{k}}\left\{ \left(\frac{-\left|\sqrt{p_{N}}\mathbf{h}_{k}\mathbf{s}_{m,j}+n_{k}\right|^{2}}{\sigma_{k}^{2}}\log_{2}e\right)\right.
\]

\begin{equation}
\left.+\log_{2}\left(1+\stackrel[i=1,i\neq j]{M^{N}}{\sum}e^{\frac{-\left|\sqrt{p_{N}}\mathbf{h}_{k}\mathbf{s}_{m,i}+n_{k}\right|^{2}+\left|\sqrt{p_{N}}\mathbf{h}_{k}\mathbf{s}_{m,j}+n_{k}\right|^{2}}{\sigma_{k}^{2}}}\right)\right\} .
\end{equation}

\noindent where $j\in\left\{ 1:M^{N}\right\} $. Please note that,
in case $j=m$, $e^{\frac{-\left|\sqrt{p_{N}}\mathbf{h}_{k}\mathbf{s}_{m,j}+n_{k}\right|^{2}}{\sigma_{k}^{2}}}=e^{\frac{-\left|n_{k}\right|^{2}}{\sigma_{k}^{2}}}$.
Finally, with the use of some auxiliary notation the second term can
be expressed as 

\begin{equation}
T_{1}=\frac{1}{M^{N}}\stackrel[m=1]{M^{N}}{\sum}\mathcal{E}_{\mathbf{h,}n_{k}}\left\{ \left(\Upsilon_{1}+\log_{2}\left(1+\Xi_{1}\right)\right)\right\} ,
\end{equation}

\noindent where $\Upsilon_{1}=\frac{-\left|\sqrt{p_{N}}\mathbf{h}_{k}\mathbf{s}_{m,j}+n_{k}\right|^{2}\log_{2}e}{\sigma_{k}^{2}}$
and $\Xi_{1}=\stackrel[i=1,i\neq j]{M^{N}}{\sum}e^{\frac{-\left|\sqrt{p_{N}}\mathbf{h}_{k}\mathbf{s}_{m,i}+n_{k}\right|^{2}+\left|\sqrt{p_{N}}\mathbf{h}_{k}\mathbf{s}_{m,j}+n_{k}\right|^{2}}{\sigma_{k}^{2}}}.$
Likewise, by following similar steps for $T_{2}$ we can get

\begin{equation}
T_{2}=\frac{1}{M^{N-1}}\stackrel[\underset{c\neq k}{c=1}]{M^{N-1}}{\sum}\mathcal{E}_{\mathbf{h,}n_{k}}\left\{ \left(\Upsilon_{2}+\log_{2}\left(1+\Xi_{2}\right)\right)\right\} ,
\end{equation}

\noindent where $\Upsilon_{2}=\frac{-\left|\sqrt{p_{N}}\mathbf{h}_{k}\mathbf{s}_{c,j}+n_{k}\right|^{2}\log_{2}e}{\sigma_{k}^{2}}$
and $\Xi_{2}=\stackrel[t=1,t\neq k,j]{M^{N-1}}{\sum}e^{\frac{-\left|\sqrt{p_{N}}\mathbf{h}_{k}\mathbf{s}_{c,t}+n_{k}\right|^{2}+\left|\sqrt{p_{N}}\mathbf{h}_{k}\mathbf{s}_{c,j}+n_{k}\right|^{2}}{\sigma_{k}^{2}}}.$
\begin{thm}
The total sum rate upper-bound of the un-precoded downlink transmission
scheme in MU-MIMO systems under PSK signaling can be calculated by

\begin{equation}
R=\stackrel[k=1]{K}{\sum}\bar{R_{k}},\label{eq:24}
\end{equation}

\noindent where $\bar{R_{k}}$ is given by (\ref{eq:22}), shown at
the top of next page.

\begin{figure*}
\begin{eqnarray}
\bar{R_{k}} & =N & \log_{2}M-\frac{1}{M^{N}}\stackrel[m=1]{M^{N}}{\sum}\left(-\log_{2}e\right.\nonumber \\
 &  & \left.+\left\{ \log_{2}\left(1+\stackrel[i=1,i\neq j]{M^{N}}{\sum}\frac{2\sigma_{k}^{2}}{2\sigma_{k}^{2}+p_{N}d_{k}^{-m}\lambda_{mi}}\right)\right\} \right)+\frac{1}{M^{N-1}}\stackrel[\underset{c\neq k}{c=1}]{M^{N-1}}{\sum}\left(-\log_{2}e\right.\nonumber \\
 &  & \left.+\left\{ \log_{2}\left(1+\stackrel[t=1,t\neq k,j]{M^{N-1}}{\sum}\frac{2\sigma_{k}^{2}}{2\sigma_{k}^{2}+p_{N}d_{k}^{-m}\lambda_{ct}}\right)\right\} \right).\label{eq:22}
\end{eqnarray}

\selectlanguage{american}%
\begin{centering}
\rule[0.5ex]{2.03\columnwidth}{0.8pt}
\par\end{centering}

\selectlanguage{english}%
\end{figure*}
\end{thm}
\begin{IEEEproof}
In order to derive the average rate in this scenario, we need to find
the averages of $T_{1}$ and $T_{2}$ over the noise and the channel
states. To start with, the average of the first term in $T_{1}$,
$\mathcal{E}_{\mathbf{h,}n_{k}}\left\{ \Upsilon_{1}\right\} $, can
be obtained as

\[
\mathcal{E}_{\mathbf{h,}n_{k}}\left\{ \frac{-\left|\sqrt{p_{N}}\mathbf{h}_{k}\mathbf{s}_{m,j}+n_{k}\right|^{2}}{\sigma_{k}^{2}}\log_{2}e\right\} 
\]

\[
=\mathcal{E}_{\mathbf{h,}n_{k}}\left(\left|\sqrt{p_{N}}\mathbf{h}_{k}\mathbf{s}_{m,j}\right|^{2}+\left|n_{k}\right|^{2}+2\left(\left(\sqrt{p_{N}}\mathbf{h}_{k}\mathbf{s}_{m,j}\right)^{H}n_{k}\right)\right)\frac{-\log_{2}e}{\sigma_{k}^{2}}
\]

\[
=\mathcal{E}_{\mathbf{h}}\left\{ -\left(\left|\sqrt{p_{N}}\mathbf{h}_{k}\mathbf{s}_{m,j}\right|^{2}+\sigma_{k}^{2}\right)\frac{\log_{2}e}{\sigma_{k}^{2}}\right\} ,
\]

\begin{equation}
=-\left(p\varpi_{k}\left\Vert \mathbf{s}_{m,j}\right\Vert ^{2}+\sigma_{k}^{2}\right)\frac{\log_{2}e}{\sigma_{k}^{2}}\qquad\qquad
\end{equation}

\noindent which can be reduced to $-\log_{2}e,\textrm{ }$by choosing
$j=m$. In order to calculate the average of the second term in $T_{1}$,
since log is a concave function, we can apply Jensen\textquoteright s
inequality, which implicates that $\mathcal{E}\left\{ \log\left(1+\Xi_{1}\right)\right\} \leq\log\left(1+\mathcal{E}\left\{ \Xi_{1}\right\} \right)$\cite{Twway}.
Therefore, using Jensen inequality the upper bound can be obtained
as

\[
\mathcal{E}_{\mathbf{h,}n_{k}}\left\{ \log_{2}\left(1+\stackrel[i=1,i\neq j]{M^{N}}{\sum}e^{\frac{-\left|\sqrt{p_{N}}\mathbf{h}_{k}\mathbf{s}_{m,i}+n_{k}\right|^{2}+\left|\sqrt{p_{N}}\mathbf{h}_{k}\mathbf{s}_{m,j}+n_{k}\right|^{2}}{\sigma_{k}^{2}}}\right)\right\} \triangleq
\]

\begin{equation}
\log_{2}\left(1+\stackrel[i=1,i\neq j]{M^{N}}{\sum}\mathcal{E}_{\mathbf{h,}n_{k}}\left\{ e^{\frac{-\left|\sqrt{p_{N}}\mathbf{h}_{k}\mathbf{s}_{m,i}+n_{k}\right|^{2}+\left|\sqrt{p_{N}}\mathbf{h}_{k}\mathbf{s}_{m,j}+n_{k}\right|^{2}}{\sigma_{k}^{2}}}\right\} \right).
\end{equation}

Since $n_{k}$ has Gaussian distribution, the average over the noise
can be derived as

\[
\mathcal{E}_{n_{k}}\left\{ e^{\frac{\left|\sqrt{p_{N}}\mathbf{h}_{k}\mathbf{s}_{m,j}+n_{k}\right|^{2}-\left|\sqrt{p_{N}}\mathbf{h}_{k}\mathbf{s}_{m,i}+n_{k}\right|^{2}}{\sigma_{k}^{2}}}\right\} =
\]

\begin{equation}
\frac{1}{\pi\sigma^{2}}\underset{n_{k}}{\int}e^{-\frac{\left|\sqrt{p_{N}}\mathbf{h}_{k}\mathbf{s}_{m,i}+n_{k}\right|^{2}-\left|\sqrt{p_{N}}\mathbf{h}_{k}\mathbf{s}_{m,j}+n_{k}\right|^{2}+\left|n_{k}\right|^{2}}{\sigma_{k}^{2}}}dn_{k}.\label{eq:12}
\end{equation}

Using the integrals of exponential function in \cite{book2}, we can
find

\[
\mathcal{E}_{n_{k}}\left\{ e^{\frac{\left|\sqrt{p_{N}}\mathbf{h}_{k}\mathbf{s}_{m,j}+n_{k}\right|^{2}-\left|\sqrt{p_{N}}\mathbf{h}_{k}\mathbf{s}_{m,i}+n_{k}\right|^{2}}{\sigma_{k}^{2}}}\right\} =
\]

\begin{equation}
\qquad\qquad\qquad\qquad e^{-\frac{\left|\sqrt{p_{N}}\mathbf{h}_{k}\mathbf{s}_{m,i}\right|^{2}-\left|\sqrt{p_{N}}\mathbf{h}_{k}\mathbf{s}_{m,j}\right|^{2}}{2\sigma_{k}^{2}}}.\label{eq:13}
\end{equation}

Therefore, the average over $\mathbf{h}$ can be written as, 

\[
\mathcal{E}_{\mathbf{h}}\left\{ \mathcal{E}_{n_{k}}\left\{ e^{\frac{\left|\sqrt{p_{N}}\mathbf{h}_{k}\mathbf{s}_{m,j}+n_{k}\right|^{2}-\left|\sqrt{p_{N}}\mathbf{h}_{k}\mathbf{s}_{m,i}+n_{k}\right|^{2}}{\sigma_{k}^{2}}}\right\} \right\} =
\]

\begin{equation}
\mathcal{E}_{\mathbf{h}}\left\{ e^{-\frac{\left|\sqrt{p_{N}}\mathbf{h}_{k}\mathbf{s}_{m,i}\right|^{2}-\left|\sqrt{p_{N}}\mathbf{h}_{k}\mathbf{s}_{m,j}\right|^{2}}{2\sigma_{k}^{2}}}\right\} .\label{eq:10}
\end{equation}

Now, to derive the average over $\mathbf{h}$, it is more convenient
to use the Quadratic form as follows

\[
\varPhi_{i}=\left|\sqrt{p_{N}}\mathbf{h}_{k}\mathbf{s}_{m,i}\right|^{2}-\left|\sqrt{p_{N}}\mathbf{h}_{k}\mathbf{s}_{m,j}\right|^{2}
\]

\[
=p_{N}\mathbf{h}_{k}\underset{\mathbf{S}_{mi}}{\underbrace{\left(\mathbf{s}_{m,i}\mathbf{s}_{m,i}^{H}-\mathbf{s}_{m,j}\mathbf{s}_{m,j}^{H}\right)}}\mathbf{h}_{k}^{H},
\]

\[
=p_{N}\mathbf{h}_{k}\mathbf{S}_{mi}\mathbf{h}_{k}^{H}\qquad\qquad\qquad\qquad
\]

\begin{equation}
=\stackrel[n=1]{l}{\sum}p_{N}d_{k}^{-m}\lambda_{mi,n}\left|\mathbf{q}_{in}^{T}\mathbf{h}_{1k}\right|^{2},
\end{equation}

\noindent where $\lambda_{mi,n}$ is the $n^{th}$ eigenvalue of matrix
\textbf{$\mathbf{S}_{mi}$} and $\mathbf{q}_{in}^{T}$ is the corresponding
eigenvector. The distribution of $\varPhi_{i}$ depends on the number
of the eigenvalues $\left(l\right)$ and the values of the eigenvalues.
In case $l=1$, $\varPhi_{i}$ has exponential distribution, in case,
$l>1$ $\varPhi_{i}$ has sum of exponential distributions, and in
case all of the eigenvalues are ones and zeros, $\varPhi_{i}$ has
Gamma distribution. By choosing $j=m$ the matrix \textbf{$\mathbf{S}_{mi}$}
will have one eigenvalue $\lambda_{mi}$. Therefore, $\varPhi_{i}$
will follow exponential distribution, and we can get,

\begin{eqnarray}
\mathcal{E}_{\mathbf{\mathbf{h}}}\left\{ e^{-\frac{\varPhi_{i}}{2\sigma_{k}^{2}}}\right\}  & = & \stackrel[0]{\infty}{\int}e^{-\frac{\varPhi}{2\sigma_{k}^{2}}}\, e^{-\varPhi}d\varPhi,\nonumber \\
 & = & \frac{2\sigma_{k}^{2}}{2\sigma_{k}^{2}+d_{k}^{-m}p_{N}\lambda_{mi}}.
\end{eqnarray}

Similarly, the average of the first term in $T_{2}$, $\mathcal{E}_{\mathbf{h,}n_{k}}\left\{ \Upsilon_{2}\right\} $,
can be obtained as

\[
\mathcal{E}_{\mathbf{h,}n_{k}}\left\{ \frac{-\left|\sqrt{p_{N}}\mathbf{h}_{k}\mathbf{s}_{c,j}+n_{k}\right|^{2}}{\sigma_{k}^{2}}\log_{2}e\right\} 
\]

\begin{equation}
=-\left(p\varpi_{k}\left\Vert \mathbf{s}_{c,j}\right\Vert ^{2}+\sigma_{k}^{2}\right)\frac{\log_{2}e}{\sigma_{k}^{2}}.
\end{equation}

\noindent which can be reduced to $-\log_{2}e,\textrm{ }$by choosing
$c=j$. In order to derive the average of the second term in $T_{2}$,
using Jensen\textquoteright s inequality, $\mathcal{E}\left\{ \log\left(1+\Xi_{2}\right)\right\} \leq\log\left(1+\mathcal{E}\left\{ \Xi_{2}\right\} \right)$,
we can calculate the upper bound as

\[
\mathcal{E}_{\mathbf{h,}n_{k}}\left\{ \log_{2}\left(1+\stackrel[t=1,t\neq k,j]{M^{N-1}}{\sum}e^{\frac{-\left|\sqrt{p_{N}}\mathbf{h}_{k}\mathbf{s}_{c,t}+n_{k}\right|^{2}+\left|\sqrt{p_{N}}\mathbf{h}_{k}\mathbf{s}_{c,j}+n_{k}\right|^{2}}{\sigma_{k}^{2}}}\right)\right\} \triangleq
\]

\begin{equation}
\log_{2}\left(1+\stackrel[t=1,t\neq k,j]{M^{N-1}}{\sum}\mathcal{E}_{\mathbf{h,}n_{k}}\left\{ e^{\frac{-\left|\sqrt{p_{N}}\mathbf{h}_{k}\mathbf{s}_{c,t}+n_{k}\right|^{2}+\left|\sqrt{p_{N}}\mathbf{h}_{k}\mathbf{s}_{c,j}+n_{k}\right|^{2}}{\sigma_{k}^{2}}}\right\} \right).
\end{equation}

Since $n_{k}$ has Gaussian distribution, following similar steps
as in (\ref{eq:12}) and (\ref{eq:13}), the average over the noise
can be obtained as

\[
\mathcal{E}_{n_{k}}\left\{ e^{\frac{-\left|\sqrt{p_{N}}\mathbf{h}_{k}\mathbf{s}_{c,t}+n_{k}\right|^{2}+\left|\sqrt{p_{N}}\mathbf{h}_{k}\mathbf{s}_{c,j}+n_{k}\right|^{2}}{\sigma_{k}^{2}}}\right\} =
\]

\begin{equation}
\qquad\qquad\qquad\qquad e^{-\frac{\left|\sqrt{p_{N}}\mathbf{h}_{k}\mathbf{s}_{c,t}\right|^{2}-\left|\sqrt{p_{N}}\mathbf{h}_{k}\mathbf{s}_{c,j}\right|^{2}}{2\sigma_{k}^{2}}}.
\end{equation}

Now, the average over $\mathbf{h}$ can be written as, 

\[
\mathcal{E}_{\mathbf{h}}\left\{ \mathcal{E}_{n_{k}}\left\{ e^{\frac{-\left|\sqrt{p_{N}}\mathbf{h}_{k}\mathbf{s}_{c,t}+n_{k}\right|^{2}+\left|\sqrt{p_{N}}\mathbf{h}_{k}\mathbf{s}_{c,j}+n_{k}\right|^{2}}{\sigma_{k}^{2}}}\right\} \right\} =
\]

\begin{equation}
\mathcal{E}_{\mathbf{h}}\left\{ e^{-\frac{\left|\sqrt{p_{N}}\mathbf{h}_{k}\mathbf{s}_{c,t}\right|^{2}-\left|\sqrt{p_{N}}\mathbf{h}_{k}\mathbf{s}_{c,j}\right|^{2}}{2\sigma_{k}^{2}}}\right\} .\label{eq:10-1}
\end{equation}

In order to derive the average over $\mathbf{h}$, the Quadratic form
can be used as follows

\[
\varPhi_{t}=\left|\sqrt{p_{N}}\mathbf{h}_{k}\mathbf{s}_{c,t}\right|^{2}-\left|\sqrt{p_{N}}\mathbf{h}_{k}\mathbf{s}_{c,j}\right|^{2}
\]

\[
=p_{N}\mathbf{h}_{k}\underset{\mathbf{S}_{ct}}{\underbrace{\left(\mathbf{s}_{c,t}\mathbf{s}_{c,t}^{H}-\mathbf{s}_{c,j}\mathbf{s}_{c,j}^{H}\right)}}\mathbf{h}_{k}^{H},
\]

\[
=p_{N}\mathbf{h}_{k}\mathbf{S}_{ct}\mathbf{h}_{k}^{H}\qquad\qquad\qquad\qquad
\]

\begin{equation}
=\stackrel[n=1]{l}{\sum}p_{N}d_{k}^{-m}\lambda_{ct,n}\left|\mathbf{q}_{tn}^{T}\mathbf{h}_{1k}\right|^{2}.
\end{equation}

\noindent where $\lambda_{ct,n}$ is the $n^{th}$ eigenvalue of matrix
\textbf{$\mathbf{S}_{ct}$}and $\mathbf{q}_{tn}^{T}$ is the corresponding
eigenvector. The distribution of $\varPhi_{t}$ depends on the number
of the eigenvalues $\left(l\right)$ and the values of the eigenvalues.
By choosing $j=c$ the matrix \textbf{$\mathbf{S}_{ct}$} will have
one eigenvalue $\lambda_{ct}$, and then $\varPhi_{t}$ will follow
exponential distribution. Therefore we can get,

\begin{equation}
\mathcal{E}_{\mathbf{\mathbf{h}}}\left\{ e^{-\frac{\varPhi_{t}}{2\sigma_{k}^{2}}}\right\} =\frac{2\sigma_{k}^{2}}{2\sigma_{k}^{2}+p_{N}d_{k}^{-m}\lambda_{ct}}.
\end{equation}

\end{IEEEproof}
The average sum-rate with respect to each user location can be obtained
by averaging the derived sum-rate over all possible user locations.

\section{Zero Forcing Precoding\label{sec:Zero-Forcing-Precoding}}

In this case the BS has perfect CSI and ZF precoding technique is
implemented. Therefore, the precoding matrix can be written as \cite{zfprecoding,angLi},

\begin{equation}
\mathbf{W}\,=\frac{1}{\beta}\,\mathbf{H}^{H}\left(\mathbf{H}\mathbf{H}^{H}\right)^{-1},
\end{equation}

\noindent where $\beta$ is the scaling factor to meet the transmit
power constraint. Therefore, the received signal at the $k^{th}$
user can be expressed as,

\begin{eqnarray}
y_{k} & = & \beta\mathbf{h}_{k}\mathbf{H}^{H}\left(\mathbf{H}\mathbf{H}^{H}\right)^{-1}\mathbf{s}+n_{k},\nonumber \\
 & = & \beta\left[\mathbf{s}\right]_{k}+n_{k}.
\end{eqnarray}

Consequently, the rate in this scenario is given by 

\[
\bar{R_{k}}^{ZF}=N\log_{2}M-\,\log_{2}e\quad\quad\quad\quad\quad\quad\quad\quad\quad
\]

\begin{equation}
-\frac{1}{M^{N}}\stackrel[m=1]{M^{N}}{\sum}\mathcal{E}_{\mathbf{H,}n_{k}}\left\{ \log_{2}\stackrel[i=1]{M^{N}}{\sum}e^{\frac{-\left|\beta\left[\mathbf{s}_{m,i}\right]_{k}+n_{k}\right|^{2}}{\sigma_{k}^{2}}}\right\} .\label{eq:5}
\end{equation}

\noindent By taking the $j^{th}$ term $\left(e^{\frac{-\left|\beta\left[\mathbf{s}_{m,j}\right]_{k}+n_{k}\right|^{2}}{\sigma_{k}^{2}}}\right)$
out, (\ref{eq:5}) can be expressed as 

\[
\bar{R_{k}}^{ZF}=N\log_{2}M-\,\log_{2}e\quad\quad\quad\quad\quad\quad\quad\quad\quad\quad\quad\quad\quad\quad\quad
\]

\begin{equation}
-\frac{1}{M^{N}}\stackrel[m=1]{M^{N}}{\sum}\mathcal{E}_{\mathbf{H,}n_{k}}\left\{ \left(\Upsilon+\log_{2}\left(1+\Xi\right)\right)\right\} ,\label{eq:31-3}
\end{equation}

\noindent where $j\in\left[1,M^{N}\right]$ , $\Upsilon=\frac{-\left|\beta\left[\mathbf{s}_{m,j}\right]_{k}+n_{k}\right|^{2}}{\sigma_{k}^{2}}\log_{2}e$
and $\Xi=\stackrel[i=1,i\neq j]{M^{N}}{\sum}e^{\frac{-\left|\beta\left[\mathbf{s}_{m,i}\right]_{k}+n_{k}\right|^{2}+\left|\beta\left[\mathbf{s}_{m,j}\right]_{k}+n_{k}\right|^{2}}{\sigma_{k}^{2}}}.$
\begin{thm}
The total sum rate upper-bound of the ZF transmission scheme in MU-MIMO
systems under PSK signaling can be calculated by

\begin{equation}
R^{ZF}=\stackrel[k=1]{K}{\sum}\bar{R_{k}}^{ZF},\label{eq:24-1-2}
\end{equation}

\noindent where $\bar{R_{k}}^{ZF}$ is given by (\ref{eq:37-2}),
shown at the top of next page.

\begin{figure*}
\[
\bar{R_{k}}^{ZF}=N\log_{2}M-\,\log_{2}e-\frac{1}{M^{N}}\stackrel[m=1]{M^{N}}{\sum}\left\{ \left(-\left(\left|\beta\left[\mathbf{s}_{m,j}\right]_{k}\right|^{2}+\sigma_{k}^{2}\right)\frac{\log_{2}e}{\sigma_{k}^{2}}\right)\right.
\]

\begin{eqnarray}
+\left.\log_{2}\left(1+\stackrel[i=1,i\neq j]{M^{N}}{\sum}e^{-\frac{\left|\beta\left[\mathbf{s}_{m,i}\right]_{k}\right|^{2}-\left|\beta\left[\mathbf{s}_{m,j}\right]_{k}\right|^{2}}{2\sigma_{k}^{2}}}\right)\right\} \label{eq:37-2}
\end{eqnarray}

\selectlanguage{american}%
\begin{centering}
\rule[0.5ex]{2.03\columnwidth}{0.8pt}
\par\end{centering}

\selectlanguage{english}%
\end{figure*}
\end{thm}
\begin{IEEEproof}
To derive the average sum rate in this case, firstly we need to derive
the average of the $\Upsilon$ term in (\ref{eq:31-3}). The scaling
factor in this scenario is given by, $\beta=\sqrt{\frac{p}{\mathbf{s}^{H}\left(\mathbf{H}\mathbf{H}^{H}\right)^{-1}\mathbf{s}}}$\cite{zfprecoding,angLi},
where $p$ is the power transmission. For simplicity but without loss
of generality, and in order to provide fair comparison between the
considered schemes, in this paper we consider constant power scaling
factors. It was shown that, the term, $X=\frac{\mathbf{s}^{H}\Sigma^{-1}\mathbf{s}}{\mathbf{s}^{H}\left(\mathbf{H}\mathbf{H}^{H}\right)^{-1}\mathbf{s}}$,
follows Gamma distribution \cite{pdfZF,CI1}, where $\Sigma=\mathbf{D}$,
so the average of the scaling factor can be obtained as $\beta=\frac{\sqrt{\frac{p}{\mathbf{s}^{H}\Sigma^{-1}\mathbf{s}}}\Gamma\left(\frac{3}{2}-K+N\right)}{K\sqrt{K}\left(N-K\right)!}$.
Therefore the average of term $\Upsilon$ in (\ref{eq:31-3}), $\mathcal{E}_{\mathbf{H,}n_{k}}\left\{ \Upsilon\right\} $,
can be obtained as

\[
\mathcal{E}_{\mathbf{H,}n_{k}}\left\{ \frac{-\left|\beta\left[\mathbf{s}_{m,j}\right]_{k}+n_{k}\right|^{2}}{\sigma_{k}^{2}}\log_{2}e\right\} =\quad\quad\quad\quad\quad\quad\quad\quad\quad
\]

\begin{equation}
\quad\quad\left\{ -\left(\left|\beta\left[\mathbf{s}_{m,j}\right]_{k}\right|^{2}+\sigma_{k}^{2}\right)\frac{\log_{2}e}{\sigma_{k}^{2}}\right\} .
\end{equation}

\noindent Now, in order to calculate the average of the last term
in (\ref{eq:31-3}), $\mathcal{E}_{\mathbf{H,}n_{k}}\left\{ \log_{2}\left(1+\Xi\right)\right\} $,
using Jensen inequality, the upper bound can be written as

\[
\mathcal{E}_{\mathbf{H,}n_{k}}\left\{ \log_{2}\left(1+\stackrel[i=1,i\neq j]{M^{N}}{\sum}e^{\frac{-\left|\beta\left[\mathbf{s}_{m,i}\right]_{k}+n_{k}\right|^{2}+\left|\beta\left[\mathbf{s}_{m,j}\right]_{k}+n_{k}\right|^{2}}{\sigma_{k}^{2}}}\right)\right\} \triangleq
\]

\begin{equation}
\log_{2}\left(1+\stackrel[i=1,i\neq j]{M^{N}}{\sum}\mathcal{E}_{\mathbf{H,}n_{k}}\left\{ e^{\frac{-\left|\beta\left[\mathbf{s}_{m,i}\right]_{k}+n_{k}\right|^{2}+\left|\beta\left[\mathbf{s}_{m,j}\right]_{k}+n_{k}\right|^{2}}{\sigma_{k}^{2}}}\right\} \right).
\end{equation}

Since $n_{k}$ has Gaussian distribution, the average over $n_{k}$
can be derived as

\[
\mathcal{E}_{n_{k}}\left\{ e^{\frac{-\left|\beta\left[\mathbf{s}_{m,i}\right]_{k}+n_{k}\right|^{2}+\left|\beta\left[\mathbf{s}_{m,j}\right]_{k}+n_{k}\right|^{2}}{\sigma_{k}^{2}}}\right\} =
\]

\begin{equation}
\frac{1}{\pi\sigma_{k}^{2}}\underset{n_{k}}{\int}e^{-\frac{\left|\beta\left[\mathbf{s}_{m,i}\right]_{k}+n_{k}\right|^{2}-\left|\beta\left[\mathbf{s}_{m,j}\right]_{k}+n_{k}\right|^{2}+\left|n_{k}\right|^{2}}{\sigma_{k}^{2}}}dn_{k}.
\end{equation}

Using the integrals of exponential function in \cite{book2}, we can
find 

\[
\mathcal{E}_{n_{k}}\left\{ e^{\frac{-\left|\beta\left[\mathbf{s}_{m,i}\right]_{k}+n_{k}\right|^{2}+\left|\beta\left[\mathbf{s}_{m,j}\right]_{k}+n_{k}\right|^{2}}{\sigma_{k}^{2}}}\right\} =
\]

\begin{equation}
e^{-\frac{\left|\beta\left[\mathbf{s}_{m,i}\right]_{k}\right|^{2}-\left|\beta\left[\mathbf{s}_{m,j}\right]_{k}\right|^{2}}{2\sigma^{2}}}\label{eq:3-1}
\end{equation}

\end{IEEEproof}
\noindent

\section{Constructive Interference Precoding\label{sec:Constructive-Interference-Precod}}

The concept and characterization for the CI have been extensively
investigated in MIMO systems \cite{chrismag1,chrismag2,CI1}. In order
to avoid repetition and for more details, we refer the reader to the
aforementioned works in this paper. Here in this section, we analyze
the performance of CI precoding technique in MU-MIMO systems, for
the first time. We focus on the recent closed form CI precoding scheme,
where the precoding matrix is given by \cite{angLi},

\begin{equation}
\mathbf{W}\boldsymbol{s}=\frac{1}{K}\beta\,\mathbf{H}^{H}\left(\mathbf{H}\mathbf{H}^{H}\right)^{-1}\textrm{diag}\left\{ \mathbf{V}^{-1}\mathbf{u}\right\} \mathbf{s},
\end{equation}

\noindent where $\beta$ is the scaling factor to meet the transmit
power constraint, which can be expressed as \cite{angLi} $\beta=\sqrt{\frac{p}{\mathbf{u}^{H}\mathbf{V}^{-1}\mathbf{u}}},$
while $\mathbf{1}^{H}\mathbf{u}=\mathbf{1}$ and $\mathbf{V}=\textrm{diag}\left(\mathbf{s}^{H}\right)\left(\mathbf{H}\mathbf{H}^{H}\right)^{-1}\textrm{diag}\left(\mathbf{s}\right).$
For the sake of comparison, the normalization factor $\beta$ is designed
to ensure that the long-term total transmit power at the source is
constrained, and it is given by \cite{Twway} $\beta=\sqrt{\frac{p}{\mathbf{u}^{H}\mathcal{E}\left[\mathbf{V}^{-1}\right]\mathbf{u}}},$
where $\mathcal{E}\left[\mathbf{V}^{-1}\right]=\textrm{diag}\left(\mathbf{s}^{H}\right)^{-1}\mathcal{E}\left[\left(\mathbf{H}\mathbf{H}^{H}\right)\right]\left(\textrm{diag}\left(\mathbf{s}\right)\right)^{-1}=\textrm{diag}\left(\mathbf{s}^{H}\right)^{-1}\, N\mathbf{\Sigma}\left(\textrm{diag}\left(\mathbf{s}\right)\right)^{-1}$
\cite{bookaspects}. The received signal at the $k^{th}$ user now
can be written as,

\[
y_{k}=\frac{\beta}{K}\mathbf{h}_{k}\,\mathbf{H}^{H}\left(\mathbf{H}\mathbf{H}^{H}\right)^{-1}\textrm{diag}\left\{ \mathbf{V}^{-1}\mathbf{u}\right\} \mathbf{s}+n_{k},\quad
\]

\begin{equation}
=\frac{\beta}{K}\mathbf{a}_{k}\left(\textrm{diag}\left(\mathbf{s}^{H}\right)\right)^{-1}\mathbf{H}\mathbf{H}^{H}\left(\textrm{diag}\left(\mathbf{s}\right)\right)^{-1}\mathbf{u}\left[\boldsymbol{s}\right]_{k}+n_{k}
\end{equation}

\noindent where $\mathbf{a}_{k}$ is a $1\times K$ vector all the
elements of this vector are zeros except the $k^{th}$ element is
one. Following the principles of CI, and the symmetric properties
of the PSK constellation, the rate at the user $k$ can be written
as,

\[
\bar{R_{k}}^{CI}=N\log_{2}M-\,\log_{2}e\quad\quad\quad\quad\quad\quad\quad\quad\quad
\]

\begin{equation}
-\frac{1}{M^{N}}\stackrel[m=1]{M^{N}}{\sum}\mathcal{E}_{\mathbf{H,}n_{k}}\left\{ \left(\Upsilon+\log_{2}\left(1+\Xi\right)\right)\right\} ,\label{eq:31-1}
\end{equation}

\noindent wherewhere $\Upsilon=\frac{-\left|\mathbf{h}_{k}\mathbf{\mathbf{W}}\boldsymbol{s}_{m,j}+n_{k}\right|^{2}\log_{2}e}{\sigma_{k}^{2}}$
and 

\noindent $\Xi=\stackrel[i=1,i\neq j]{M^{N}}{\sum}e^{\frac{-\left|\mathbf{h}_{k}\mathbf{\mathbf{W}}\boldsymbol{s}_{m,i}+n_{k}\right|^{2}+\left|\mathbf{h}_{k}\mathbf{\mathbf{W}}\boldsymbol{s}_{m,j}+n_{k}\right|^{2}}{\sigma_{k}^{2}}}.$
\begin{thm}
The total sum rate upper-bound of the CI transmission scheme in MU-MIMO
systems under PSK signaling can be calculated by

\begin{equation}
R^{CI}=\stackrel[k=1]{K}{\sum}\bar{R_{k}}^{CI},\label{eq:24-1-1}
\end{equation}

\noindent where 

\begin{eqnarray}
\bar{R_{k}}^{CI} & = & N\,\log_{2}M-\,\log_{2}e-\frac{1}{M^{N}}\stackrel[m=1]{M^{N}}{\sum}\left(-\log_{2}e\right.\nonumber \\
 &  & \left.+\left\{ \log_{2}\left(1+\stackrel[i=1,i\neq j]{M^{N}}{\sum}\Lambda_{m,i}\right)\right\} \right).\label{eq:49}
\end{eqnarray}

\noindent and $\Lambda_{m,i}$ is given by (\ref{eq:39}).
\end{thm}
\begin{figure*}
\[
\Lambda_{m,i}=\left(\frac{2^{\left(\frac{1}{2}\left(N-K-1\right)\right)}K^{\left(N-K+1\right)}\left|\left[\boldsymbol{s}_{m,i}\right]_{k}\right|^{-2+K-N}}{\left(N-K\right)!}\right)\left(\left(\frac{c_{k}^{2}}{\sigma_{k}^{2}}\right)^{\frac{1}{2}\left(K-N-1\right)}\right)
\]

\[
\times\left(\left(c_{k}^{2}\left|\left[\boldsymbol{s}_{m,i}\right]_{k}\right|\right)\Gamma\left(\frac{1}{2}\left(N-K+1\right)\right)\textrm{\ensuremath{_{1}\textrm{F}_{1}}}\left(\frac{1}{2}\left(N-K+1\right),\frac{1}{2},\frac{K^{2}\sigma_{k}^{2}}{2c_{k}^{2}\left|\left[\boldsymbol{s}_{m,i}\right]_{k}\right|^{2}}\right)\right.
\]

\begin{equation}
\left.-\sqrt{2}K\, c_{k}\,\sigma_{k}\Gamma\left(\frac{1}{2}\left(N-K+2\right)\right){}_{1}\textrm{F}_{1}\left(\frac{1}{2}\left(N-K+2\right),\frac{3}{2},\frac{K^{2}\sigma_{k}^{2}}{2c_{k}^{2}\left|\left[\boldsymbol{s}_{m,i}\right]_{k}\right|^{2}}\right)\right).\label{eq:39}
\end{equation}

\selectlanguage{american}%
\centering{}\rule[0.5ex]{2.03\columnwidth}{0.8pt}\selectlanguage{english}%
\end{figure*}

\begin{IEEEproof}
To start with, the average of the $\Upsilon$ term in (\ref{eq:31-1}),
$\mathcal{E}_{\mathbf{H,}n_{k}}\left\{ \Upsilon\right\} $ , can be
obtained as

\[
\mathcal{E}_{\mathbf{H,}n_{k}}\left\{ \frac{-\left|\frac{\beta}{K}\left[\,\textrm{diag}\left\{ \mathbf{V}^{-1}\mathbf{u}\right\} \boldsymbol{s}_{m,j}\right]_{k}+n_{k}\right|^{2}}{\sigma_{k}^{2}}\log_{2}e\right\} =
\]

\begin{equation}
\mathcal{E}_{\mathbf{H}}\left\{ -\left(\left|\frac{\beta}{K}\mathbf{a}_{k}\textrm{diag}\left\{ \mathbf{V}^{-1}\mathbf{u}\right\} \left[\boldsymbol{s}_{m,j}\right]_{k}\right|^{2}+\sigma_{k}^{2}\right)\frac{\log_{2}e}{\sigma_{k}^{2}}\right\} .\label{eq:42}
\end{equation}

\noindent In \cite{bookaspects} it was shown that, the term $Z=\frac{\mathbf{a}_{k}\left(\textrm{diag}\left(\mathbf{s}^{H}\right)\right)^{-1}\left(\mathbf{H}\mathbf{H}^{H}\right)\left(\textrm{diag}\left(\mathbf{s}\right)\right)^{-1}\mathbf{u}\left[\boldsymbol{s}_{m,j}\right]_{k}}{\mathbf{a}_{k}\left(\textrm{diag}\left(\mathbf{s}^{H}\right)\right)^{-1}\left(\mathbf{\Sigma}\right)\left(\textrm{diag}\left(\mathbf{s}\right)\right)^{-1}\mathbf{u}\left[\boldsymbol{s}_{m,j}\right]_{k}}$
has Gamma distribution with $N$ degrees of freedom. Hence the average
in (\ref{eq:42}) can be derived as in (\ref{eq:18}), shown at the
top of next page. 

\begin{figure*}
\[
\mathcal{E}_{\mathbf{H}}\left\{ -\left(\left|\frac{\beta}{K}\mathbf{a}_{k}\textrm{diag}\left\{ \mathbf{V}^{-1}\mathbf{u}\right\} \left[\boldsymbol{s}_{m,j}\right]_{k}\right|^{2}+\sigma_{k}^{2}\right)\frac{\log_{2}e}{\sigma_{k}^{2}}\right\} \quad\quad\quad\quad\quad\quad\quad\quad\quad\quad
\]

\[
=-\left(\mathcal{E}_{\mathbf{H}}\left\{ \left|\frac{\beta\left(\mathbf{a}_{k}\left(\textrm{diag}\left(\mathbf{s}^{H}\right)\right)^{-1}\left(\mathbf{\Sigma}\right)\left(\textrm{diag}\left(\mathbf{s}\right)\right)^{-1}\mathbf{u}\left[\boldsymbol{s}_{m,j}\right]_{k}\right)}{K}Z\right|^{2}\right\} +\sigma_{k}^{2}\right)\frac{\log_{2}e}{\sigma_{k}^{2}}
\]

\begin{equation}
\quad\quad\quad\quad=\left(\frac{-\Gamma\left(2+N\right)}{\sigma_{k}^{2}\Gamma\left(N\right)}\left|\frac{\beta\left(\mathbf{a}_{k}\left(\textrm{diag}\left(\mathbf{s}^{H}\right)\right)^{-1}\left(\mathbf{\Sigma}\right)\left(\textrm{diag}\left(\mathbf{s}\right)\right)^{-1}\mathbf{u}\left[\boldsymbol{s}_{m,j}\right]_{k}\right)}{K}\right|^{2}-1\right)\log_{2}e.\label{eq:18}
\end{equation}

\selectlanguage{american}%
\centering{}\rule[0.5ex]{2.03\columnwidth}{0.8pt}\selectlanguage{english}%
\end{figure*}

\noindent In order to calculate the average of the last term in (\ref{eq:31-1}),
$\mathcal{E}_{\mathbf{H,}n_{k}}\left\{ \log_{2}\left(1+\Xi\right)\right\} $,
using Jensen inequality, the upper bound can be calculated as follows.
Since $n_{k}$ has Gaussian distribution, the average over the noise
can be obtained as

\[
\mathcal{E}_{n_{k}}\left\{ e^{\frac{-\left|\mathbf{h}_{k}\mathbf{\mathbf{W}}\boldsymbol{s}_{m,i}+n_{k}\right|^{2}+\left|\mathbf{h}_{k}\mathbf{\mathbf{W}}\boldsymbol{s}_{m,j}+n_{k}\right|^{2}}{\sigma_{k}^{2}}}\right\} 
\]

\begin{equation}
=e^{-\frac{\left|\frac{\beta}{K}\mathbf{a}_{k}\textrm{diag}\left\{ \mathbf{V}^{-1}\mathbf{u}\right\} \boldsymbol{s}_{m,i}\right|^{2}-\left|\frac{\beta}{K}\mathbf{a}_{k}\textrm{diag}\left\{ \mathbf{V}^{-1}\mathbf{u}\right\} \boldsymbol{s}_{m,j}\right|^{2}}{2\sigma_{k}^{2}}}.\label{eq:19}
\end{equation}

Now, the average of (\ref{eq:19}) over $\mathbf{H}$ can be written
as in (\ref{eq:36-1}), shown at the top of next page, 
\begin{figure*}
\[
\mathcal{E}_{\mathbf{H}}\left\{ e^{-\frac{\left|\frac{\beta}{K}\mathbf{a}_{k}\textrm{diag}\left\{ \mathbf{V}^{-1}\mathbf{u}\right\} \boldsymbol{s}_{m,i}\right|^{2}-\left|\frac{\beta}{K}\mathbf{a}_{k}\textrm{diag}\left\{ \mathbf{V}^{-1}\mathbf{u}\right\} \boldsymbol{s}_{m,j}\right|^{2}}{2\sigma_{k}^{2}}}\right\} \qquad\qquad\qquad\qquad\qquad\qquad\qquad\qquad
\]

\[
=\mathcal{E}_{\mathbf{H}}\left\{ e^{-\frac{\left|\frac{\beta}{K}\mathbf{a}_{k}\left(\textrm{diag}\left(\mathbf{s}^{H}\right)\right)^{-1}\left(\mathbf{H}\mathbf{H}^{H}\right)\left(\textrm{diag}\left(\mathbf{s}\right)\right)^{-1}\mathbf{u}\left[\boldsymbol{s}_{m,i}\right]_{k}\right|^{2}-\left|\frac{\beta}{K}\,\mathbf{a}_{k}\left(\textrm{diag}\left(\mathbf{s}^{H}\right)\right)^{-1}\left(\mathbf{H}\mathbf{H}^{H}\right)\left(\textrm{diag}\left(\mathbf{s}\right)\right)^{-1}\mathbf{u}\left[\boldsymbol{s}_{m,j}\right]_{k}\right|^{2}}{2\sigma_{k}^{2}}}\right\} ,
\]

\[
=\mathcal{E}_{\mathbf{H}}\left\{ e^{-\frac{\left|\frac{\beta\left(\mathbf{a}_{k}\left(\textrm{diag}\left(\mathbf{s}^{H}\right)\right)^{-1}\Sigma\left(\textrm{diag}\left(\mathbf{s}\right)\right)^{-1}\mathbf{u}\right)}{K}\, X\left[\boldsymbol{s}_{m,i}\right]_{k}\right|^{2}-\left|\frac{\beta\left(\mathbf{a}_{k}\left(\textrm{diag}\left(\mathbf{s}^{H}\right)\right)^{-1}\Sigma\left(\textrm{diag}\left(\mathbf{s}\right)\right)^{-1}\mathbf{u}\right)}{K}\, X\left[\boldsymbol{s}_{m,j}\right]_{k}\right|^{2}}{2\sigma_{k}^{2}}}\right\} ,
\]

\begin{equation}
=\mathcal{E}_{\mathbf{H}}\left\{ e^{-\frac{\chi_{i,j}}{2\sigma_{k}^{2}}}\right\} .\qquad\qquad\qquad\qquad\qquad\qquad\qquad\qquad\qquad\qquad\qquad\qquad\qquad\qquad\label{eq:36-1}
\end{equation}

\selectlanguage{american}%
\centering{}\rule[0.5ex]{2.03\columnwidth}{0.8pt}\selectlanguage{english}%
\end{figure*}
 where $\chi_{i,j}=\left|\frac{\beta\left(\mathbf{a}_{k}\left(\textrm{diag}\left(\mathbf{s}^{H}\right)\right)^{-1}\Sigma\left(\textrm{diag}\left(\mathbf{s}\right)\right)^{-1}\mathbf{u}\right)}{K}\, X\left[\boldsymbol{s}_{m,i}\right]_{k}\right|^{2}-\left|\frac{\beta\left(\mathbf{a}_{k}\left(\textrm{diag}\left(\mathbf{s}^{H}\right)\right)^{-1}\Sigma\left(\textrm{diag}\left(\mathbf{s}\right)\right)^{-1}\mathbf{u}\right)}{K}\, X\left[\boldsymbol{s}_{m,j}\right]_{k}\right|^{2}$,
$X=\frac{\mathbf{a}_{k}\left(\textrm{diag}\left(\mathbf{s}^{H}\right)\right)^{-1}\left(\mathbf{H}\mathbf{H}^{H}\right)\left(\textrm{diag}\left(\mathbf{s}\right)\right)^{-1}\mathbf{u}}{\mathbf{a}_{k}\left(\textrm{diag}\left(\mathbf{s}^{H}\right)\right)^{-1}\Sigma\left(\textrm{diag}\left(\mathbf{s}\right)\right)^{-1}\mathbf{u}}$
.

\noindent Since $\chi_{i,j}$ has Gamma distribution \cite{pdfZF,CI1};
we can find the average as in (\ref{eq:38}), 
\begin{figure*}
\[
\mathcal{E}_{\mathbf{\mathbf{H}}}\left\{ e^{-\frac{\chi_{i,j}}{2\sigma^{2}}}\right\} =\Lambda_{i,j}=\left(\frac{2^{\left(\frac{1}{2}\left(N-K-1\right)\right)}K^{\left(N-K+1\right)}}{\left(N-K\right)!}\right)\left(\left(\frac{c_{k}^{2}\xi}{\sigma_{k}^{2}}\right)^{\frac{1}{2}\left(K-N-1\right)}\right)
\]

\[
\quad\quad\quad\quad\quad\quad\quad\quad\times\left(\left(c_{k}^{2}\left(\left|\left[\boldsymbol{s}_{m,i}\right]_{k}\right|^{2}\right)\right)\Gamma\left(\frac{1}{2}\left(N-K+1\right)\right)\textrm{\ensuremath{_{1}\textrm{F}_{1}}}\left(\frac{1}{2}\left(N-K+1\right),\frac{1}{2},\frac{K^{2}\sigma_{k}^{2}}{2c_{k}^{2}\xi}\right)\right.
\]

\[
\quad\quad\quad\quad\quad\quad\quad-\left(c_{k}^{2}\left(\left|\left[\boldsymbol{s}_{m,j}\right]_{k}\right|^{2}\right)\right)\Gamma\left(\frac{1}{2}\left(N-K+1\right)\right)\textrm{\ensuremath{_{1}\textrm{F}_{1}}}\left(\frac{1}{2}\left(N-K+1\right),\frac{1}{2},\frac{K^{2}\sigma_{k}^{2}}{2c_{k}^{2}\xi}\right)
\]

\begin{equation}
\left.\quad\quad\quad\quad\quad\quad\quad\quad\quad\quad\quad-\sqrt{2}K\,\sigma_{k}^{2}\sqrt{\frac{\xi}{\sigma_{k}^{2}}}\Gamma\left(\frac{1}{2}\left(N-K+2\right)\right){}_{1}\textrm{F}_{1}\left(\frac{1}{2}\left(N-K+2\right),\frac{3}{2},\frac{K^{2}\sigma_{k}^{2}}{2c_{k}^{2}\xi}\right)\right).\label{eq:38}
\end{equation}

\selectlanguage{american}%
\centering{}\rule[0.5ex]{2.03\columnwidth}{0.8pt}\selectlanguage{english}%
\end{figure*}
 where $\xi=\left(\left|\left[\boldsymbol{s}_{m,i}\right]_{k}\right|^{2}-\left|\left[\boldsymbol{s}_{m,j}\right]_{k}\right|^{2}\right)$,
$c_{k}=\frac{\beta\left(\mathbf{a}_{k}\left(\textrm{diag}\left(\mathbf{s}^{H}\right)\right)^{-1}\Sigma\left(\textrm{diag}\left(\mathbf{s}\right)\right)^{-1}\mathbf{u}\right)}{K}$
and $\textrm{\ensuremath{_{1}\textrm{F}_{1}}}$ is the Hypergeometric
function. By choosing $j=m$, $\chi_{i,j}$ reduces to, $\chi_{i}=\left|\frac{\beta\left(\mathbf{a}_{k}\left(\textrm{diag}\left(\mathbf{s}^{H}\right)\right)^{-1}\Sigma\left(\textrm{diag}\left(\mathbf{s}\right)\right)^{-1}\mathbf{u}\right)}{K}\, X\left[\boldsymbol{s}_{m,i}\right]_{k}\right|^{2}$
and the expression can be further simplified as in (\ref{eq:39}).
\end{IEEEproof}

\section{Fairness for MU-MIMO Systems with PSK Signaling\label{sec:Transmission-Power-Minimization}}

In this section we consider the fairness among the users for the system
model under consideration. Firstly, based on the derived expressions
in the previous sections, we calculate the minimum power required
to achieve a target data rate, and then we propose a fairness algorithm
that can be used to provide fairness in MU-MIMO systems with finite
alphabet signals.

\subsection{Minimum Power Transmission}

Consider that the BS transmits the messages at a target data rate.
Let the target data rate be denoted by $R_{T}$, so that for perfect
transmission the achievable rate at the $k^{th}$ user, $\bar{R_{k}}$,
has to satisfy the condition, $\bar{R_{k}}\geq R_{T}$ . In order
to determine the minimum power transmission required to achieve $R_{T}$,
we simplify the derived expressions in the previous sections as follows.

\subsubsection{un-precoded downlink Transmission}

To find the minimum power transmission in this case we simplify (\ref{eq:22})
into,

$\,$

\[
R_{T}=\bar{R_{k}}\quad\quad\quad\quad\quad\quad\quad\quad\quad\quad\quad\quad\quad\quad\quad\quad\quad\quad\quad\quad\quad\quad\quad\quad\quad\quad\quad
\]

\[
=X-\left\{ \frac{1}{M^{N}}\stackrel[m=1]{M^{N}}{\sum}\log_{2}\left(1+\stackrel[i=1,i\neq j]{M^{N}}{\sum}\frac{2\sigma_{k}^{2}}{2\sigma_{k}^{2}+p_{k}d_{k}^{-m}\lambda_{mi}}\right)\right\} 
\]

\begin{equation}
+\left\{ \frac{1}{M^{N-1}}\stackrel[\underset{c\neq k}{c=1}]{M^{N-1}}{\sum}\log_{2}\left(1+\stackrel[t=1,t\neq k]{M^{N-1}}{\sum}\frac{2\sigma_{k}^{2}}{2\sigma_{k}^{2}+p_{k}d_{k}^{-m}\lambda_{ct}}\right)\right\} ,
\end{equation}

\noindent where $X=N\,\log_{2}M$. The minimum $p_{k}$ can be obtained
by solving,

\[
X-\left\{ \frac{1}{M^{N}}\stackrel[m=1]{M^{N}}{\sum}\log_{2}\left(1+\stackrel[i=1,i\neq j]{M^{N}}{\sum}\frac{2\sigma_{k}^{2}}{2\sigma_{k}^{2}+p_{k}d_{k}^{-m}\lambda_{mi}}\right)\right\} +
\]

\begin{equation}
\frac{1}{M^{N-1}}\stackrel[\underset{c\neq k}{c=1}]{M^{N-1}}{\sum}\log_{2}\left(1+\stackrel[\underset{t\neq k}{t=1}]{M^{N-1}}{\sum}\frac{2\sigma_{k}^{2}}{2\sigma_{k}^{2}+p_{k}d_{k}^{-m}\lambda_{ct}}\right)-R_{T}=0\label{eq:50}
\end{equation}

Therefore, minimum value of $p_{k}$ is the value that satisfies (\ref{eq:50}).
By exploiting the symmetric properties of the constellation \cite[Eq(5)]{MIMOPSK1},
(\ref{eq:50}) can be reduced to 

\[
X-\left\{ \log_{2}\left(1+\stackrel[i=1,i\neq j]{M^{N}}{\sum}\frac{2\sigma_{k}^{2}}{2\sigma_{k}^{2}+p_{k}d_{k}^{-m}\lambda_{\bar{m}i}}\right)\right\} 
\]

\begin{equation}
+\left\{ \log_{2}\left(1+\stackrel[t=1,t\neq k]{M^{N-1}}{\sum}\frac{2\sigma_{k}^{2}}{2\sigma_{k}^{2}+p_{k}d_{k}^{-m}\lambda_{\bar{c}t}}\right)\right\} -R_{T}=0,\label{eq:50-1}
\end{equation}

\noindent where $\bar{m}$ and $\bar{c}=\left[1,1,...1\right]^{T}$\cite{MIMOPSK1}.
Hence, (\ref{eq:50-1}) can be written as 

\begin{equation}
\frac{1+\stackrel[t=1,t\neq k]{M^{N-1}}{\sum}\frac{2\sigma_{k}^{2}}{2\sigma_{k}^{2}+p_{k}d_{k}^{-m}\lambda_{\bar{c}t}}}{1+\stackrel[i=1,i\neq j]{M^{N}}{\sum}\frac{2\sigma_{k}^{2}}{2\sigma_{k}^{2}+p_{k}d_{k}^{-m}\lambda_{\bar{m}i}}}=Y,
\end{equation}

and 

\[
\stackrel[t=1,t\neq k]{M^{N-1}}{\sum}\frac{2}{2+\frac{p_{k}}{\sigma_{k}^{2}}d_{k}^{-m}\lambda_{\bar{c}t}}
\]

\begin{equation}
-Y\stackrel[i=1,i\neq j]{M^{N}}{\sum}\frac{2}{2+\frac{p_{k}}{\sigma_{k}^{2}}d_{k}^{-m}\lambda_{\bar{m}i}}=Y-1,
\end{equation}

\noindent where $Y=2^{R_{T}-X}$, therefore at high SNR, the minimum
value of the transmission power $p_{k}$ can be expressed as

\begin{equation}
p_{k}=\frac{\stackrel[t=1,t\neq k]{M^{N-1}}{\sum}\frac{2\sigma_{k}^{2}}{d_{k}^{-m}\lambda_{\bar{c}t}}-Y\stackrel[i=1,i\neq j]{M^{N}}{\sum}\frac{2\sigma_{k}^{2}}{d_{k}^{-m}\lambda_{\bar{m}i}}}{Y-1}.\label{eq:power1}
\end{equation}

\subsubsection{Zero Forcing Precoding}

To find the minimum power transmission in ZF precoding case, we simplify
(\ref{eq:37-2}) into,

\[
R_{T}=\bar{R_{k}}^{ZF}\quad\quad\quad\quad\quad\quad\quad\quad\quad\quad\quad\quad\quad\quad\quad\quad\quad
\]

\[
=X-\frac{1}{M^{N}}\stackrel[m=1]{M^{N}}{\sum}\left\{ \left(-\left(\left|\beta\left[\mathbf{s}_{m,j}\right]_{k}\right|^{2}+\sigma_{k}^{2}\right)\frac{\log_{2}e}{\sigma_{k}^{2}}\right)\right.\qquad
\]

\begin{equation}
+\left.\log_{2}\left(1+\stackrel[i=1,i\neq j]{M^{N}}{\sum}e^{-\frac{\left|\beta\left[\mathbf{s}_{m,i}\right]_{k}\right|^{2}-\left|\beta\left[\mathbf{s}_{m,j}\right]_{k}\right|^{2}}{2\sigma_{k}^{2}}}\right)\right\} 
\end{equation}

\noindent where $X=N\log_{2}M-\,\log_{2}e$. The minimum $p_{k}$
can be obtained by solving,

\[
X-\frac{1}{M^{N}}\stackrel[m=1]{M^{N}}{\sum}\left\{ \left(-\left(\left|\beta\left[\mathbf{s}_{m,j}\right]_{k}\right|^{2}+\sigma_{k}^{2}\right)\frac{\log_{2}e}{\sigma_{k}^{2}}\right)\right.
\]

\begin{equation}
+\left.\log_{2}\left(1+\stackrel[i=1,i\neq j]{M^{N}}{\sum}e^{-\frac{\left|\beta\left[\mathbf{s}_{m,i}\right]_{k}\right|^{2}-\left|\beta\left[\mathbf{s}_{m,j}\right]_{k}\right|^{2}}{2\sigma_{k}^{2}}}\right)\right\} -R_{T}=0\label{eq:55}
\end{equation}

The minimum $p_{k}$ is the value that satisfies (\ref{eq:55}). Again
using the simplification in \cite[Eq(5)]{MIMOPSK1}, (\ref{eq:55})
reduces to 

\[
X-\left\{ \left(-\left(\left|\beta\left[\mathbf{s}_{\bar{m},j}\right]_{k}\right|^{2}+\sigma_{k}^{2}\right)\frac{\log_{2}e}{\sigma_{k}^{2}}\right)\right.
\]

\begin{equation}
+\left.\log_{2}\left(1+\stackrel[i=1,i\neq j]{M^{N}}{\sum}e^{-\frac{\left|\beta\left[\mathbf{s}_{\bar{m},i}\right]_{k}\right|^{2}-\left|\beta\left[\mathbf{s}_{\bar{m},j}\right]_{k}\right|^{2}}{2\sigma_{k}^{2}}}\right)\right\} -R_{T}=0
\end{equation}

which can be written as

\[
\left\{ \left(-\left(\left|\beta\left[\mathbf{s}_{\bar{m},j}\right]_{k}\right|^{2}+\sigma_{k}^{2}\right)\frac{\log_{2}e}{\sigma_{k}^{2}}\right)\right.
\]

\begin{equation}
+\left.\log_{2}\left(1+\stackrel[i=1,i\neq j]{M^{N}}{\sum}e^{-\frac{\left|\beta\left[\mathbf{s}_{\bar{m},i}\right]_{k}\right|^{2}-\left|\beta\left[\mathbf{s}_{\bar{m},j}\right]_{k}\right|^{2}}{2\sigma_{k}^{2}}}\right)\right\} =X-R_{T}
\end{equation}

Therefore, in high SNR the minimum transmission power can be obtained
by

\begin{equation}
p_{k}=\frac{\sigma_{k}^{2}\left(R_{T}-N\,\log_{2}M\right)}{\log_{2}e\,\zeta^{2}\varpi_{k}^{2}\left|\left[\mathbf{s}_{\bar{m},j}\right]_{k}\right|^{2}}\label{eq:power2-1}
\end{equation}

\noindent where $\zeta=\frac{\Gamma\left(\frac{3}{2}-K+N\right)}{\sqrt{\left(\mathbf{s}^{H}\Sigma^{-1}\mathbf{s}\right)}K\sqrt{K}\left(N-K\right)!}.$

\subsubsection{Constructive Interference Precoding}

To find the minimum transmission power in CI precoding scheme we simplify
(\ref{eq:49}) into,

\begin{eqnarray}
R_{T} & = & \bar{R_{k}}^{CI}\nonumber \\
 & = & X-\left\{ \frac{1}{M^{N}}\stackrel[m=1]{M^{N}}{\sum}\log_{2}\left(1+\stackrel[i=1,i\neq j]{M^{N}}{\sum}\Lambda_{m,i}\right)\right\} ,\\
\nonumber 
\end{eqnarray}

\noindent where $X=N\,\log_{2}M$. The minimum $p_{k}$ can be obtained
by solving,

\begin{equation}
X-\left\{ \frac{1}{M^{N}}\stackrel[m=1]{M^{N}}{\sum}\log_{2}\left(1+\stackrel[i=1,i\neq j]{M^{N}}{\sum}\Lambda_{m,i}\right)\right\} -R_{T}=0.\label{eq:65-2}
\end{equation}

Therefore, the minimum value of $p_{k}$ is the value that satisfies
(\ref{eq:65-2}). Using the simplification in \cite[Eq(5)]{MIMOPSK1},
(\ref{eq:65-2}) becomes, 

\begin{equation}
X-\left\{ \log_{2}\left(1+\stackrel[i=1,i\neq j]{M^{N}}{\sum}\Lambda_{\bar{m},i}\right)\right\} -R_{T}=0,\label{eq:65-2-1}
\end{equation}

\noindent  which can be written as, 

\begin{equation}
\stackrel[i=1,i\neq j]{M^{N}}{\sum}\Lambda_{\bar{m},i}=2^{X-R_{T}}-1.
\end{equation}

The Hypergeometric function is defined as,

\begin{equation}
\textrm{\ensuremath{_{1}\textrm{F}_{1}}}\left(a,b,z\right)=\stackrel[\upsilon=0]{\infty}{\sum}\frac{\left(a\right)_{\upsilon}}{\left(b\right)_{\upsilon}}\frac{z^{\upsilon}}{\upsilon!},\label{eq:67}
\end{equation}

\noindent where $\left(a\right)_{\upsilon}$ and $\left(b\right)_{\upsilon}$
are Pochhammer symbols. By substituting (\ref{eq:67}) into (\ref{eq:39})
we can notice that, at high SNR or when number of the users is much
smaller than number of the antennas, only the first term in (\ref{eq:67})
has great impact on the value of Hypergeometric function, and therefore
the other terms can be ignored with high accuracy. Hence, at high
SNR we can get

\begin{equation}
\stackrel[i=1,i\neq j]{M^{N}}{\sum}\left(AC_{1}C_{2}\right)=2^{X-R_{T}}-1,\label{eq:72}
\end{equation}

\noindent where $A=\left(\frac{2^{\left(\frac{1}{2}\left(N-K-1\right)\right)}K^{\left(N-K+1\right)}\left|\left[\boldsymbol{s}_{\bar{m},i}\right]_{k}\right|^{-2+K-N}}{\left(N-K\right)!}\right)$,
$C_{1}=\left(\left(\frac{c_{k}^{2}}{\sigma_{k}^{2}}\right)^{\frac{1}{2}\left(K-N-1\right)}\right),$ 

\noindent $C_{2}=\left(\left(c_{k}^{2}\left|\left[\boldsymbol{s}_{\bar{m},i}\right]_{k}\right|\right)\Gamma\left(\frac{1}{2}\left(N-K+1\right)\right)\right).$
From (\ref{eq:72}) we can find, 

\begin{equation}
a_{1}\stackrel[i=1,i\neq j]{M^{N}}{\sum}\left|\left[\boldsymbol{s}_{\bar{m},i}\right]_{k}\right|^{-1+K-N}\left(\left(c_{k}^{2}\right)^{\frac{1}{2}\left(K-N+1\right)}\right)=\left(2^{X-R_{T}}-1\right),\label{eq:73}
\end{equation}

\begin{equation}
a_{1}\stackrel[i=1,i\neq j]{M^{N}}{\sum}\left|\left[\boldsymbol{s}_{\bar{m},i}\right]_{k}\right|^{-1+K-N}\left(\left(\frac{\beta}{a_{2}}\right)^{\left(K-N+1\right)}\right)=\left(2^{X-R_{T}}-1\right),\label{eq:73-1}
\end{equation}

\noindent where $a_{1k}=\frac{\sigma_{k}^{\left(N-K+1\right)}2^{\left(\frac{1}{2}\left(N-K-1\right)\right)}K^{\left(N-K+1\right)}\Gamma\left(\frac{1}{2}\left(N-K+1\right)\right)}{\left(N-K\right)!}$
and

\noindent $a_{2k}=\frac{K}{\left(\mathbf{a}_{k}\left(\textrm{diag}\left(\mathbf{s}^{H}\right)\right)^{-1}\Sigma\left(\textrm{diag}\left(\mathbf{s}\right)\right)^{-1}\mathbf{u}\right)}$.
 Finally, the minimum transmission power in CI scenario can be obtained
by

\begin{equation}
p_{k}=\left(\left(\sqrt[\left(\frac{K-N+1}{2}\right)]{\frac{\left(2^{X-R_{T}}-1\right)}{a_{1k}\stackrel[i=1,i\neq j]{M^{N}}{\sum}\left(a_{2k}a_{3}\right)^{N-K-1}\left|\left[\boldsymbol{s}_{\bar{m},i}\right]_{k}\right|^{-1+K-N}}}\right)\right).\label{eq:power3}
\end{equation}

\noindent where $a_{3}=\sqrt{\mathbf{u}^{H}\textrm{diag}\left(\mathbf{s}^{H}\right)^{-1}\, N\mathbf{\Sigma}\left(\textrm{diag}\left(\mathbf{s}\right)\right)^{-1}\mathbf{u}}$.

\noindent

\subsection{$Max-Min$ Fairness Algorithm}

In this section, based on the achievable data rate, the fairness problem
is formulated. Specifically, we propose a power allocation scheme
which maximizes the minimum user rate, whilst satisfying the total
power constraint as in the following expression,

\begin{eqnarray}
\underset{p_{k}}{\max}\,\,\,\,\,\,\underset{k=1,...K}{\min}\,\,\,\,\, R_{k}\nonumber \\
\textrm{s.t.}\stackrel[k=1]{K}{\sum}p_{k}\leq P_{t}\:\:\:\:\:\:\:\:\:\:\label{eq:78}
\end{eqnarray}

\noindent where $P_{t}$ is the total power. As we can see from the
previous sections that, the achievable data rate expression, $R_{k}$,
is complex and this complexity makes the optimization problem in (\ref{eq:78})
hard to solve using standard optimization solvers. However, some iterative
algorithms can be used to solve a power allocation problem.  Consequently,
for the target data rate $R_{T}$, we can consider the following problem,

\begin{eqnarray}
\underset{p_{k}}{\max}\,\, R_{T}\qquad\qquad\nonumber \\
\textrm{s.t. }R_{k}\geqslant R_{T},\: k=1,..,K\nonumber \\
\stackrel[k=1]{K}{\sum}p_{k}\leq P_{t},\, p_{k}\neq0\,\,\label{eq:79}
\end{eqnarray}

\begin{algorithm*}
\begin{enumerate}
\item \noindent Initialize $R_{TLB}=0,\textrm{ and }R_{TUB}=\log_{2}M.$
\item While $\left(R_{TUB}-R_{TLB}\geq\epsilon\right)$do
\item \noindent Set $R_{T}=\frac{R_{TLB}+R_{TUB}}{2}.$
\item \noindent Obtain $p_{1},...,p_{K}$ from (\ref{eq:power1}), (\ref{eq:power2-1})
and (\ref{eq:power3}).
\item \noindent If $\left(\stackrel[k=1]{K}{\sum}p_{k}\leq P_{t}\right)$then, 
\item \noindent Set $R_{TLB}=R_{T}$ ; $R^{*}=R_{T}$
\item \noindent Else 
\item \noindent Set $R_{TUB}=R_{T}$
\end{enumerate}
\protect\caption{Optimal Algorithm for $R^{*}$.}
\end{algorithm*}

According to the last formula in (\ref{eq:79}), the optimal objective
function value of (\ref{eq:78}) $\left(R^{*}\right)$is larger than
or equal to $R_{T}$. It has been presented in literature that, the
rate in such systems is an increasing function with the power, and
there is minimum power value, $p_{m}$, in which the rate reaches
its maximum value; as the power increases beyond this amount the rate
will be constant. Based on this fact, the required power for each
user, $p_{1},...,p_{K}$, in each transmission scheme can be calculated
using the derived equations (\ref{eq:power1}), (\ref{eq:power2-1})
and (\ref{eq:power3}), and the optimal $R_{T}$ can be obtained using
Bisection method as explained in Algorithm 1, shown at the top of
next page.

\section{Numerical Results\textcolor{red}{\label{sec:Numerical-Results}}}

In this section some numerical results of the considered transmission
techniques are presented. Monte-Carlo simulations are conducted, in
which channel coefficients are randomly generated in each simulation
run. Assuming the BS transmission power is $p$, and the users have
same noise power $\sigma^{2}$, the SNR ratio can be defined as SNR
= $\frac{p}{\sigma^{2}}$, when the channels are normalized and the
path loss exponent is chosen to be $m=2.7$. 
\begin{figure}
\noindent \begin{centering}
\subfloat[\label{fig:1a}Rate versus SNR with different types of input, when
$d_{1}=d_{2}=1m$]{\begin{centering}
\includegraphics[scale=0.6]{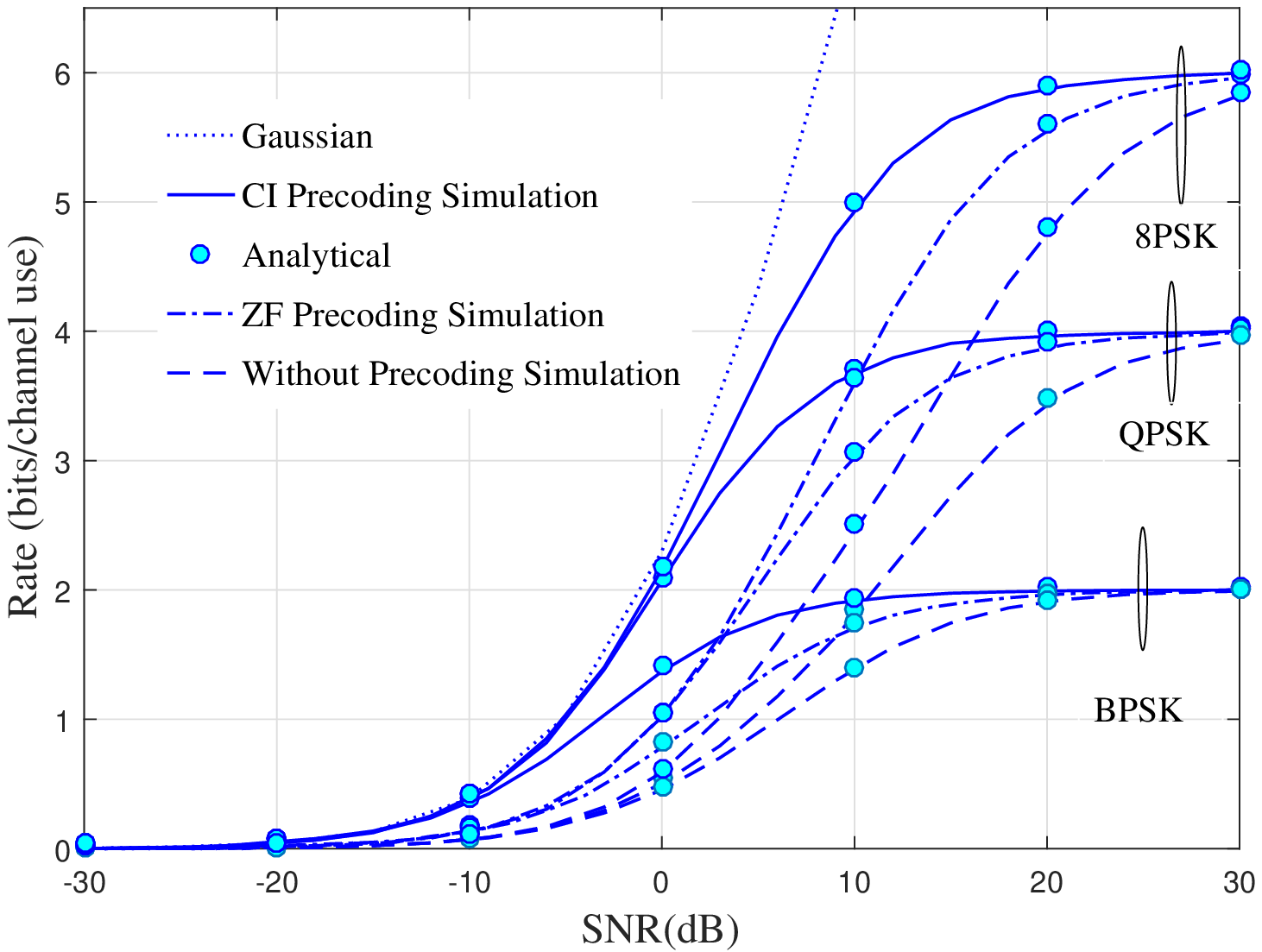}
\par\end{centering}

}
\par\end{centering}

\noindent \begin{centering}
\subfloat[\label{fig:1b}Rate versus SNR with different types of input, when
the users are randomly distributed.]{\noindent \begin{centering}
\includegraphics[scale=0.6]{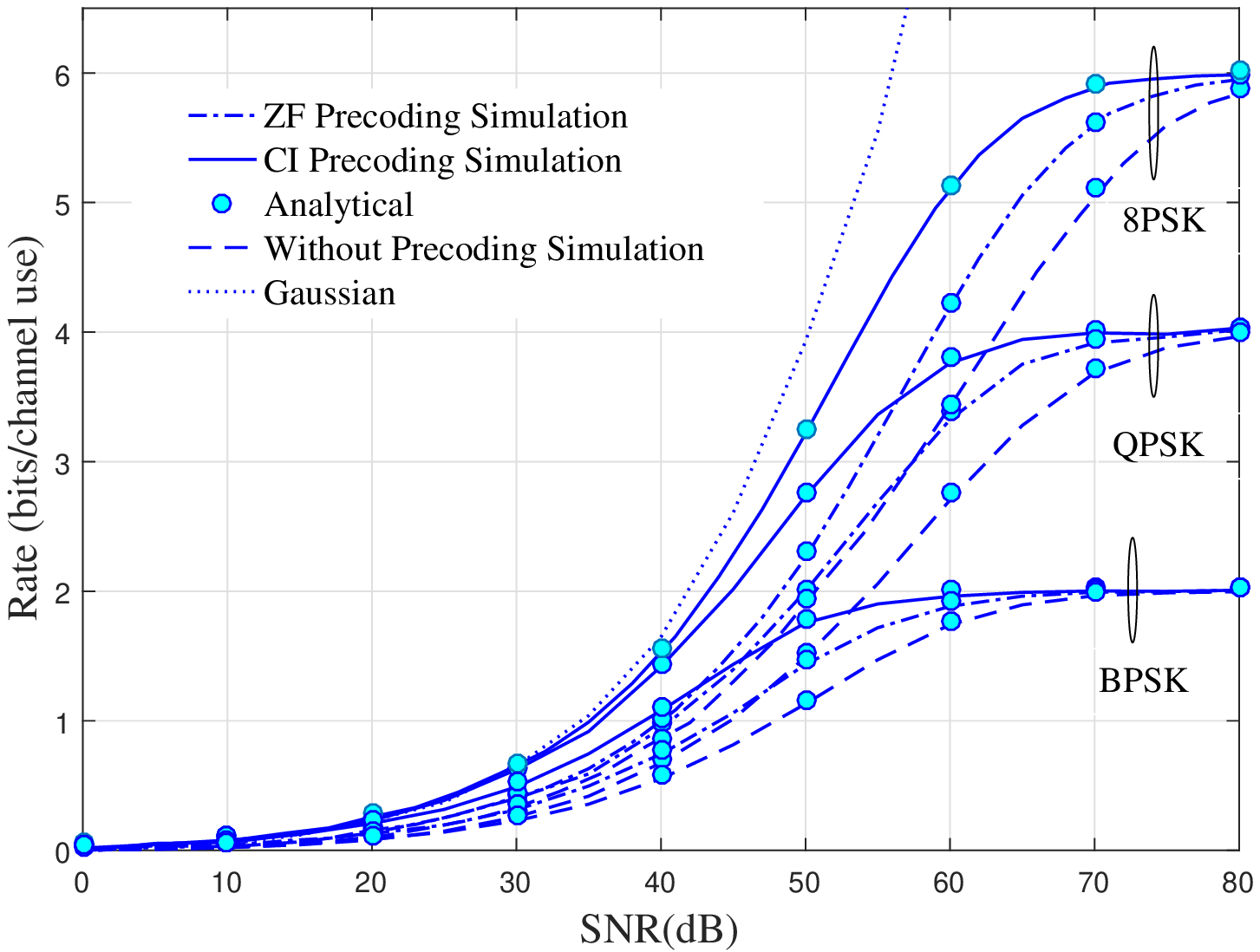}
\par\end{centering}

}
\par\end{centering}

\protect\caption{\label{fig:1} Rate versus SNR with different types of input, when
$N=2,K=2$.}
\end{figure}
\begin{figure*}
\noindent \begin{centering}
\subfloat[\label{fig:2a}Sum rate versus SNR with different types of input,
when $d_{1}=d_{2}=d_{3}=1m$.]{\noindent \begin{centering}
\includegraphics[scale=0.6]{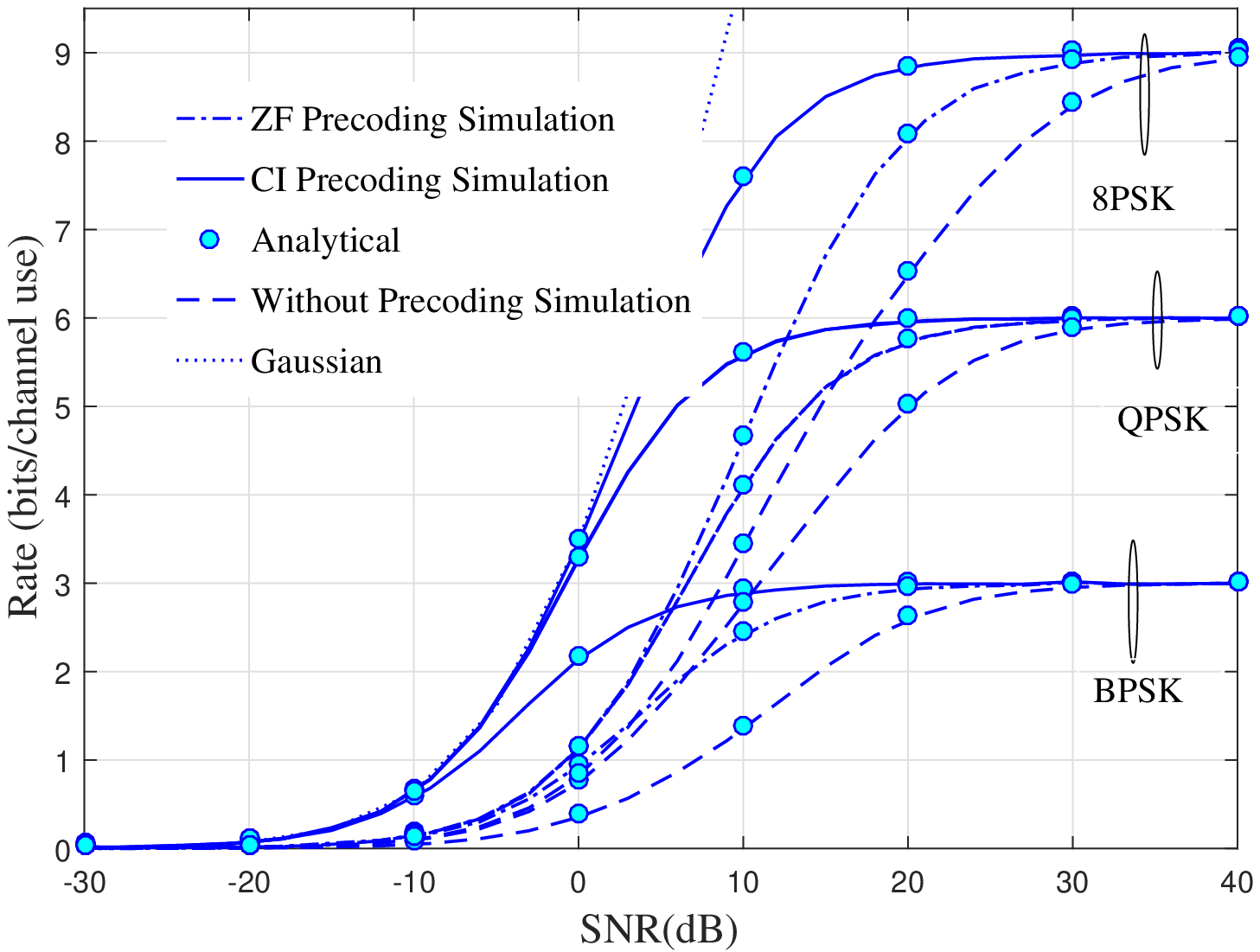}
\par\end{centering}

}\subfloat[\label{fig:2b}Sum rate versus SNR with different types of input,
when the users are randomly distributed.]{\noindent \begin{centering}
\includegraphics[scale=0.6]{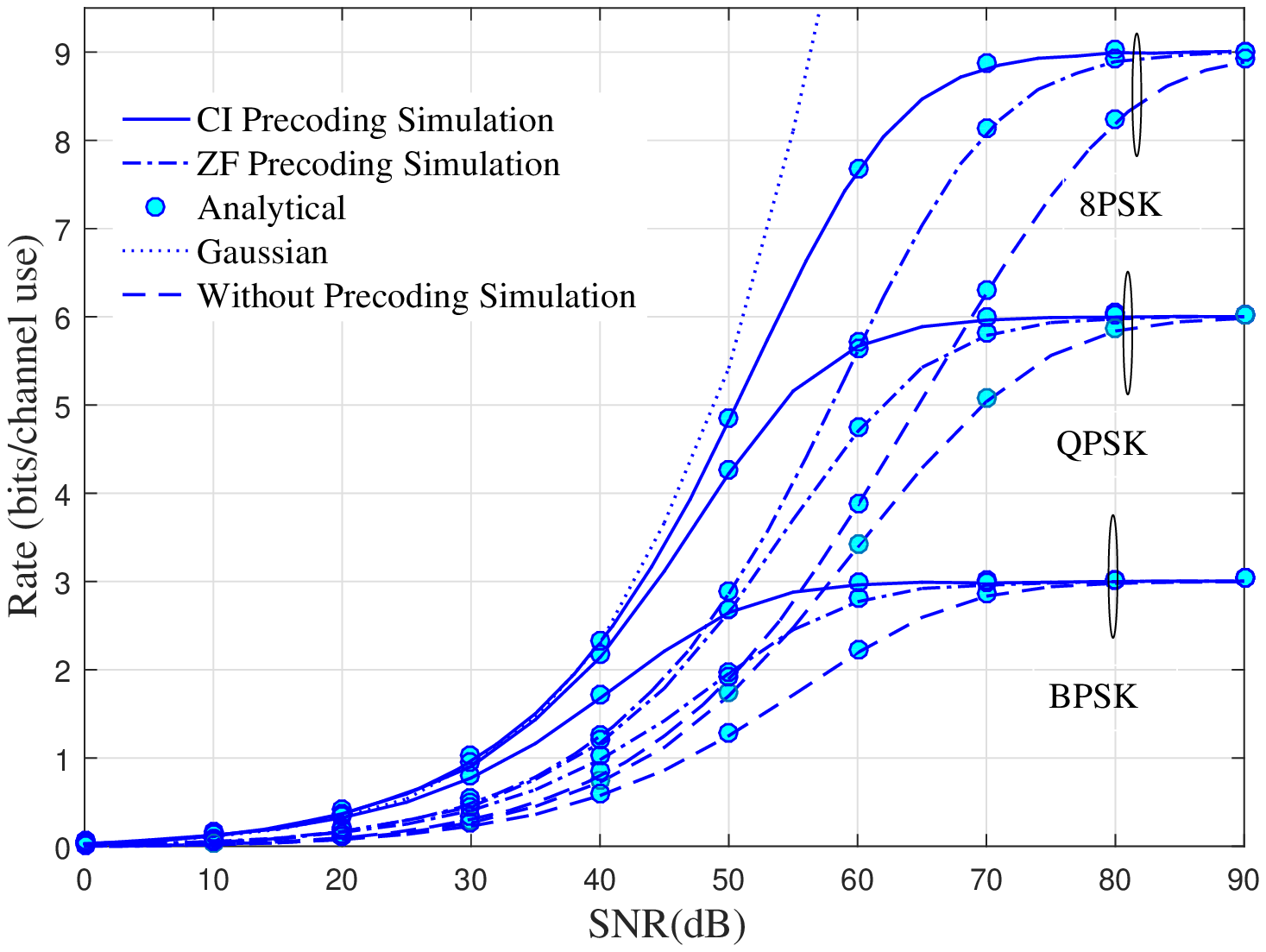}
\par\end{centering}

}
\par\end{centering}

\protect\caption{\label{fig:2}Rate versus SNR with different types of input, when
$N=3,K=3$.}
\end{figure*}

Fig. \ref{fig:1} illustrates the sum-rate for the three transmission
schemes, subject to different types of input, BPSK, QPSK and 8PSK,
when $N=2$, and $K=2$. Fig. \ref{fig:1a}, presents the sum-rate
when the distances between the BS and the users are normalized to
unit value, .i.e, without the impact of the path-loss. Fig. \ref{fig:1b}
shows the sum-rate when the users are uniformly distributed inside
a circle area with a radius of 80m, and no user is closer to the BS
than 10m where the BS is located at the center of this area. The good
agreement between the analytical and simulated results confirms the
validity of our analysis in the previous sections. From this figure,
we have several observations. Firstly, it is evident that the sum
rate saturates to the value of, $K\log_{2}M$, past a certain SNR,
owing to the finite constellation; the sum rates saturate at 2 bits/s/Hz
in BPSK, at 4 bits/s/Hz in QPSK and at 6 bits/s/Hz in 8PSK. In addition,
the CI technique always outperforms the ZF and non-precoding techniques
for a wide SNR range with an up to 5dB gain in the SNR for a given
sum rate. Finally, comparing Fig. \ref{fig:1a} and Fig. \ref{fig:1b},
one can notice that, in general, increasing the distance always degrades
the achievable sum rates. In addition, when the distance between the
BS and the users increases the rate saturation occurs at high SNR
values, due to larger path-loss. Furthermore, it is worth noting that,
the gain attained by CI over the ZF does not depend on the users\textquoteright{}
locations.

To capture the influence of number of BS antennas and number of users
on the system performance, we present in Fig. \ref{fig:2} the sum-rate
for the considered transmission schemes for BPSK, QPSK and 8PSK, when
$N=3$, and $K=3$. Comparing the results in this figure with the
ones in Fig. \ref{fig:1}, it is clear that increasing $N$ and/or
$K$ leads to enhance the system performance. In addition, the CI
technique always outperforms the ZF technique with an up to 7dB gain
in the SNR for a given sum rate. Furthermore, comparing the sum rate
achieved in Fig. \ref{fig:2a} and Fig. \ref{fig:2b}, we can see
similar observations as in the case when $N=K=2$.

\begin{figure*}
\noindent \begin{centering}
\includegraphics[scale=0.6]{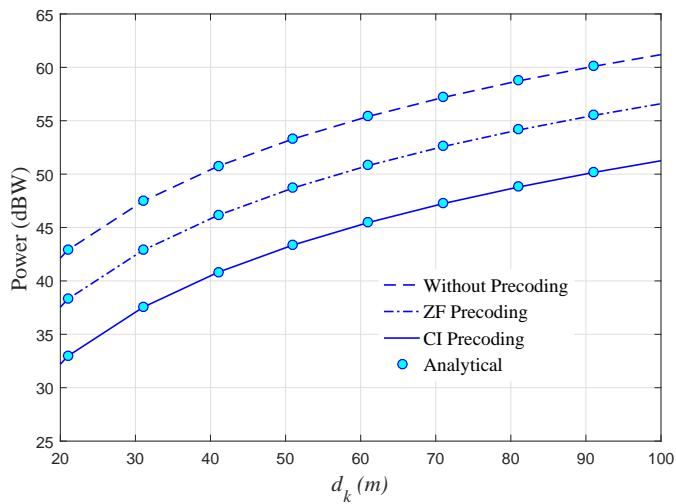}
\par\end{centering}

\protect\caption{\label{fig:4}Minimum power transmission versus $d_{k}$ with BPSK
input, when $R_{T}=0.5\,\left(\textrm{bits/s/Hz}\right),\, N=2,\textrm{and}K=2$.
}
\end{figure*}

In Fig. \ref{fig:4} we plot the minimum power transmission versus
a user distance, when $N=2$, $K=2$ and the target data rate, $R_{T}=0.5\textrm{ bits/s/Hz}$
in BPSK scenario. It should be pointed out that the results for the
conventional, ZF and CI techniques in this figure are obtained from
Section \ref{sec:Transmission-Power-Minimization}. Generally and
as anticipated, the CI technique consumes much smaller power transmission
than the other two schemes to achieve the same target data rate, and
this superiority is almost fixed with the distance. 

\begin{figure*}
\noindent \begin{centering}
\includegraphics[scale=0.6]{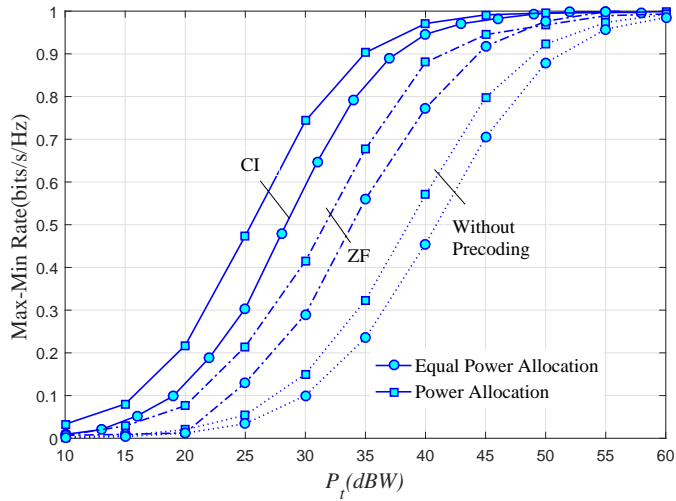}
\par\end{centering}

\protect\caption{\label{fig:Max-Min}Total max-min rate versus power with BPSK input,
when $N=2,\, K=2,\, d_{k}=15\textrm{m}.$. }
\end{figure*}

Moreover, the max-min rate of the considered system versus the total
power is shown in Fig. \ref{fig:Max-Min}. From this figure, it can
be clearly noticed that, the rate can be enhanced significantly by
using the proposed power allocation algorithm. Furthermore, the CI
always has higher sum-rate than ZF and un-precoded techniques.

\begin{figure*}
\noindent \begin{centering}
\includegraphics[scale=0.6]{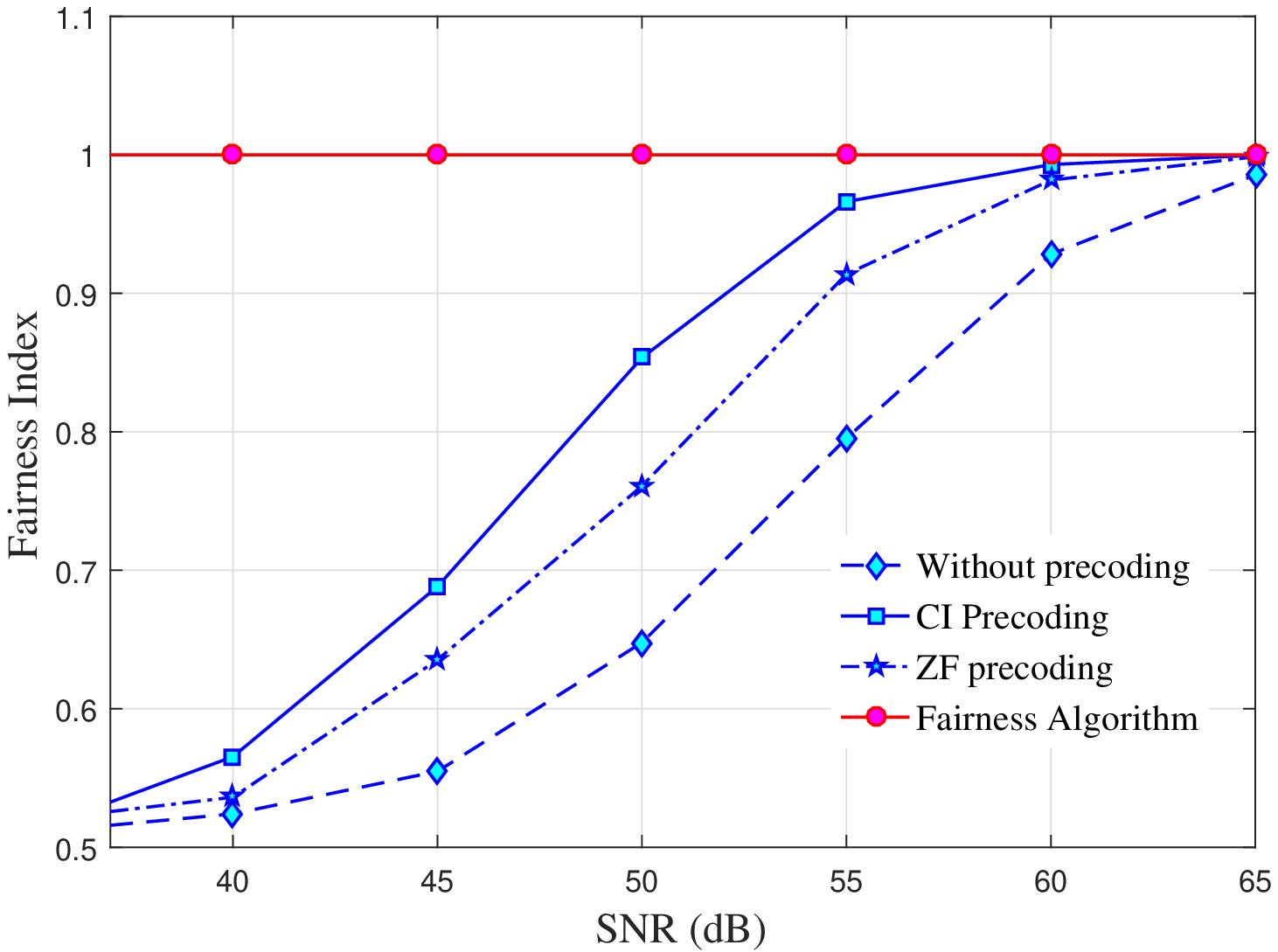}
\par\end{centering}

\protect\caption{\label{fig:Fairness-index}Fairness index versus SNR with BPSK, when
$N=\, K=2,\, d_{1}=10\textrm{m},\, d_{2}=90\textrm{m}$.}

\end{figure*}

Finally, Fig. \ref{fig:Fairness-index} illustrates the Jain\textquoteright s
fairness index versus the SNR. The fairness index is defined as \cite{newfairness},
$\frac{\left(\stackrel[k=1]{K}{\sum}R_{k}\right)^{2}}{K\stackrel[k=1]{K}{\sum}R_{k}^{2}}$,
the range of Jain\textquoteright s fairness index is between 0 and
1, where the maximum achieved when users\textquoteright{} rates are
equal. It can be observed that, in case equal power allocation transmission,
the fairness index increases as the SNR increases, and the CI achieves
higher fairness than the other transmission techniques. In addition
and as anticipated, the proposed power allocation algorithm performs
higher fairness index than equal power allocation transmission scheme.

\section{Conclusions\label{sec:Conclusions}}

In this paper we analyzed for the first time the performance of CI
precoding technique in MU-MIMO systems with a PSK input alphabet.
In light of this, new explicit analytical expressions for the average
sum rate are derived for three downlink transmission schemes: 1) without
precoding, 2) ZF precoding technique 3) CI precoding technique. In
addition, based on the derived sum-rate expressions, the minimum transmission
power that performs a target data rate was obtained for each transmission
scheme, and then power allocation algorithm has been proposed to provide
fairness among the users. The results in this work demonstrated that
no matter what the values of the system parameters are, the CI scheme
outperforms the other two schemes, and the performance gap between
the considered schemes depends essentially on the system parameters.
Furthermore, increasing the SNR enhances the sum rate to a certain
level, and increasing the distance between the BS and the users has
no impact on the gap between the minimum transmission power required
for ZF and CI to achieve same target data rate. Finally, it was shown
that, the CI can provide higher fairness than ZF technique, and the
power allocation algorithm proposed can perform high fairness index. 

\bibliographystyle{IEEEtran}
\bibliography{bib}

\end{document}